\RequirePackage{fix-cm}
\documentclass[smallextended]{svjour3}       
\smartqed  
\usepackage{cite}
\usepackage{caption}
\usepackage{subcaption}
\usepackage{graphicx, amsmath}
\usepackage{float}
\usepackage{subfloat}
\usepackage{booktabs}
\usepackage{multirow}
\graphicspath{ {./images/} }
\begin{document}

\title{Entropy correction and quasinormal modes of slowly rotating Kerr-Newman-de Sitter-like black hole in bumblebee gravity}


\author{Ningthoujam Media, Y. Priyobarta Singh, Y. Onika Laxmi \and
        T. Ibungochouba Singh* 
}


\institute{Ningthoujam Media  \at
              Department of Mathematics, Manipur University, Canchipur, 795003, India \\
              \email{medyaningthoujam@gmail.com}           
           \and
           Y. Priyobarta Singh \at
           Department of Mathematics, Manipur University, Canchipur, 795003, India \\
           \email{priyoyensh@gmail.com}
           \and
           Y. Onika Laxmi \at
           Department of Mathematics, Manipur University, Canchipur, 795003, India \\
           \email{onikalaxmi@gmail.com}
           \and
            T. Ibungochouba Singh* \at
              Department of Mathematics, Manipur University, Canchipur, 795003, India \\
               \email{ibungochouba@rediffmail.com}
}

\date{Received: date / Accepted: date}

\maketitle

\begin{abstract}
In this paper, we study the tunneling of fermion particle for slowly rotating Kerr-Newman-de Sitter-like (KNdS-like) black hole in bumblebee gravity model by applying the generalized uncertainty principle (GUP). The Hawking temperature of slowly rotating KNdS-like black hole has been modified under GUP. The quantum gravity effect reduces the rise of Hawking temperature of slowly rotating KNdS-like black hole. The modified Hawking temperature and the correction of black hole entropy are investigated by using the tunneling of fermion beyond the semiclassical approximation. We study the scalar field perturbation and effective potential of slowly rotating KNdS-like black hole in bumblebee gravity model by using Klein-G\"{o}rdon equation. The quasinormal modes for scalar perturbation is investigated using WKB approximation method and P{\"o}schl-Teller fitting method. The significant impact on the greybody factor, Hawking spectra and sparsity of Hawking radiation of the black hole are also studied in the presence of Lorentz violating parameter $L$.

\end{abstract}

\section{Introduction}
A black hole can create and emit particles as a thermal radiation from its event horizon if the quantum effects are considered. This thermal radiation emitted from black hole event horizon is known as Hawking radiation \cite{haw2}.  The gravitational potential produced by the black hole itself is encountered by Hawking radiation. As a result, the Hawking radiation is reflected and transmitted. Consequently, a blackbody spectrum is not the same as the actual spectrum seen by an asymptotic observer. The quantity which measures the variation of the Hawking radiation of black hole from a black body radiation is known as the greybody factor \cite{page1, page2}. The greybody factor can be calculated in different methods. The general rigorous bound method for transmission and reflection for one-dimensional potential scattering is one of them \cite{visser, boon1}. Other methods include matching method \cite{fernando,kim}, WKB approximation \cite{jusufi,parikh} and analytical method for the various of spin fields \cite{boon2,sakali,badawi,badawi2,gursel,kanzi1}. Another important feature of the flow of Hawking radiation is a dimensionless quantity known as its sparsity \cite{gray,hod1,hod2,miao,chow,paul}. Since the average time gap between successive radiation quanta emission is greater than the characteristic time-scale of individual Hawking emission, then the Hawking emission is known to be incredibly sparse.\\
In the theory of general relativity, black holes are among the most fascinating objects. The most significant phase in the context of general relativity for gravitational wave radiation which describes the frequencies of oscillation of the black hole is known as quasinormal modes (QNMs), which depends on the characteristics of black holes such as the charge, the mass and the spin \cite{berti1,kono1,kono2,priyo2}. The QNMs frequencies of black hole are complex numbers where the real part represents the frequency of the oscillation and its imaginary part indicates the rate of the oscillation decays \cite{det,berti2}. There are various important applications in studying their stability, horizon area, surface gravity, gravitational wave and geometrical properties of black hole. Different methods used to calculate QNMs are the WKB approximation, P{\"o}schl-Teller approximation \cite{priyo2,pos} the time-domain integration method and the continued fraction method. Cho \cite{cho} calculated the Dirac field QNMs of a Schwarzschild black hole by using the third-order of WKB approximation. Schutz and Will \cite{schutz} developed a WKB method that was expanded to higher orders in \cite{iyer}. Konoplya \cite{kono2} calculated the QNMs of a D-dimensional Schwarzschild black hole by extending the WKB approximation to sixth-order. Using the WKB approximation of sixth-order and the approximation by P{\"o}schl-Teller potential, Zhidenko \cite{zhiden} calculated low-laying QNMs of a Schwarzschild-de Sitter black hole. The QNMs of the near extremal Schwarzschild-de Sitter black hole was investigated by using P{\"o}schl-Teller approximation \cite{cardo1} and it was equal to be exactly in the near extreme regime \cite{cardo2}. One may calculate the quasinormal modes of the Kerr black hole by using the Teukolsky equations \cite{detweiler,leaver}. The QNMs was also studied numerically in the extremal Schwarzschild-de Sitter spacetime \cite{yoshida}. It is worthwhile to mention that the most significant QNMs are the lowest one with less imaginary part on the astrophysical aspect and the most significant spacetime is asymptotically flat and it will be the asymptotically de Sitter space which is supported by the latest observation data \cite{chang}.  \\
The Hawking radiation as a semiclassical quantum tunneling phenomenon where a particle moves a dynamical geometry was proposed by Parikh and Wilczek \cite{parikh}. The tunneling of semiclassical method by using Hamilton-Jacobi method was also discussed in \cite{van,medov,sri,hemm}. Applying the generalized uncertainty principle (GUP), the quantum gravity effects were studied in different black holes. It is noted that the particle tunneling rate alters from pure thermality which satisfies the unitary theory. Moreover, the minimal observable length on the Planck scale is predicted by string theory, loop quantum gravity and quantum geometry \cite{ame,nou}, which tends to the formation of GUP as \cite{adler,hao,fai,gim,murei,ablu,kesh,priyo}
\begin{eqnarray*}
\Delta x \Delta p\geq\frac{\hbar}{2}[1+\beta (\Delta p)^2],
\end{eqnarray*}
where $\beta=\frac{\alpha_0}{M_p^2}$. Here $\alpha_0$ and $M_p=\sqrt{\frac{\hbar c}{G}}$ are the dimensionless parameter and Planck mass respectively. Recently many papers based on the effects of GUP in the Hawking radiation of black holes have been studied in the literatures \cite{taw,noza,gan}. To investigate the quantum corrections to the Hawking radiation of the black hole \cite{kono2,ali}, the WKB approximation \cite{pad} is used to amalgamate the GUP with the considered wave equation.\\
In modern physics, the Lorentz dispersion relation is a fundamental relation which is closely related to both general relativity and quantum field theory. Many researchers showed that the Lorentz dispersion relationship in high energy field should be modified by Planck scale $\hbar$ in the study of string theory and quantum gravity theory. As a result, some scientists studied many interesting results considering the modified Lorentz dispersion relationship from flat spacetime to curve spacetime. Both flat spacetime and curved spacetime, many researchers investigated significant corrections to the bosons and fermions dynamic equations by taking Lorentz invariance violation and Lorentz symmetry breaking effects \cite{kruglov1,jacob,kruglov2,ding1}. Wentao et al. \cite{liu8} investigated the effects with a Lorentz violating field and showed that the coupling between the Lorentz symmetry violating field and the spacetime alleviates gravity induced entanglement degradation. Ref. \cite{liu7} studied the affects of a Lorentz symmetry violating vector field on entanglement harvesting between two particle detectors in a BTZ like black hole. Ref. \cite{liu6} also explored the affects of Lorentz symmetry violation in Einstein-Bumblebee black hole within the spectral property of vector modes. Another approach to induce the spontaneous Lorentz symmetry breaking includes the consideration of Kalb-Ramond (KR) field which non-minimally couples to gravity. If the KR field attains a non-zero vacuum expectation value, it tends to the spontaneous Lorentz symmetry violation \cite{alts}. Using Lorentz violation theory, the modified Hawking temperature and entropy of black holes have been discussed in \cite{yang1,yang2,yang3,onika1,onika2,niranjan}. Einstein-bumblebee gravity \cite{kos1,maluf1} is one of the prominent and reliable theories which contain Lorentz violation. A bumblebee vector field with a non-zero vacuum expectation value may spontaneously violate Lorentz symmetry. The associated implications of the Lorentz violation in black hole physics and cosmology have been extensively studied in the papers \cite{bluhm1,maluf2,fang,uniyal,khoda,reyes}. The first black hole solution for this efficient theory called Einstein-bumblebee gravity has been derived by Casana et al. \cite{casana}. Since then, many researchers developed the spherical symmetric black hole solutions within the framework of bumblebee gravity such as the gobal monopole \cite{gullu}, the cosmological constant \cite{ovgun1}, traversable wormhole solution \cite{ovgun2} and the Einstein-Gauss-Bonnet term \cite{ding2,oli}. Ref. \cite{ding3} proposed that the exact solution of Kerr-like black hole under the framework of bumblebee gravity model was derived. Solving Einstein-bumblebee equations, Ref. \cite{jha} also attempted to derive an exact Kerr-solution. However, Refs. \cite{ding4,maluf3} proved that the solutions derived in Refs. \cite{ding3,jha} are incorrect. Later, it is noted that their findings may be taken as an approximation solution of Einstein-bumblebee field equations \cite{wen}.

The organisation of this paper is as follows: In Sec. 2, the basic properties of slowly rotating KNdS-like black hole are studied. In Sec. 3, we investigate the tunneling of fermions near the event horizon of black hole under the influence of GUP and corresponding Hawking temperature is derived. In Sec. 4, we also study the modified Hawking temperature and entropy correction of black hole beyond the semiclassical approximation of black hole. The effective potential and the greybody radiation are also discussed by using Klein-G\"{o}rdon equation in Sec. 5. In Sec. 6, we study the quasinormal modes and Hawking sparsity of the slowly rotating KNdS-like black hole. Conclusion is given in Sec. 7. (Throughout the paper, we follow the metric convention (+,-,-,-) and the geometrized units $G=c=1$).
\section{KNdS-like black hole in bumblebee gravity}
In Boyer-Lindquist coordinates $(t,r,\theta,\phi)$, the line element of KNdS-like black hole in bumblebee gravity is given by \cite{zhou}
\begin{eqnarray}\label{k1}
ds^2&=&\frac{1}{\rho^2}[\Delta-a^2(1+L) \sin^2\theta \Delta_\theta]dt^2-(1+L)\frac{\rho^2}{\Delta}dr^2-\frac{\rho^2}{\Delta_\theta}d{\theta}^2 
-\frac{\sin^2\theta}{\rho^2 \Xi^2}\Big[ (r^2+a^2(1+L))^2 \Delta_\theta \cr&& -a^2(1+L)\sin^2\theta \Delta \Big]d{\phi^2}
+\frac{2a \sin^2\theta \sqrt{1+L}}{\rho^2 \Xi}\Big[(r^2+a^2(1+L)) \Delta_\theta  -\Delta \Big]dt d{\phi},
\end{eqnarray}
where
\begin{eqnarray*}
&&\Delta=(r^2+a^2(1+L))\Big(1-\frac{1}{3}(1+L)\Lambda r^2 \Big)-2M r+\frac{2(1+L)Q^2}{2+L},\cr
&&\Delta_\theta=1+\frac{1}{3}(1+L)^2 \Lambda a^2 \cos^2\theta,\cr
&&\rho^2=r^2+(1+L)a^2\cos^2\theta,\cr
&& \Xi=1+\frac{1}{3}(1+L)^2 \Lambda a^2.
\end{eqnarray*}
The constants $M$, $L$, $Q$, $a$ and $\Lambda$ are the black hole mass, the Lorentz violating parameter, electric charge, rotational parameter and the positive cosmological constant respectively. The metric given in Eq. \eqref{k1} becomes KNdS black hole when $L$ is absent \cite{medi}.\\
The slowly rotating KNdS-like black hole in bumblebee gravity is given by
\begin{eqnarray}\label{so}
ds^2&=&\frac{\Delta_r}{r^2}dt^2-(1+L)\frac{r^2}{\Delta_r}dr^2-r^2( d{\theta}^2+\sin^2\theta d{\phi^2})
+\frac{2a \sin^2\theta \sqrt{1+L}}{r^2}(r^2 -\Delta_r)dt d{\phi}, \nonumber\\
\end{eqnarray}
where
\begin{eqnarray*}
\Delta_r=r^2-2M r-\frac{\Lambda r^4}{3}(1+L)+\frac{2(1+L)Q^2}{2+L}.
\end{eqnarray*}
 Eq. \eqref{so} becomes slowly rotating KNdS black hole when $L$ tends to zero. If $Q=0$ and $\Lambda=0$, it represents the slowly rotating Kerr-like black hole \cite{kan}.
\begin{figure}[!htbp]
\centering
\includegraphics[width=175pt,height=155pt]{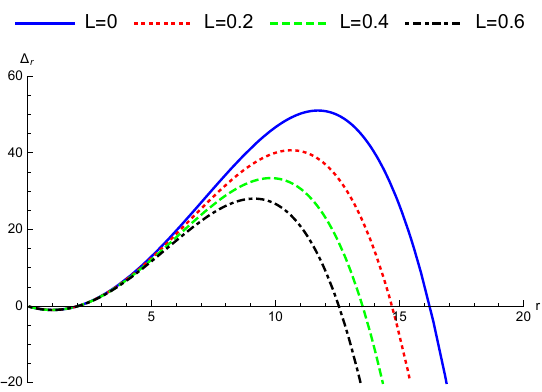}
\caption{Plot of $\Delta_r$ versus $r$ for different values of $L$ with $Q=0.1, \Lambda=0.01$ and $M=1$.}
\label{delta}
\end{figure}
The electromagnetic potential is given by
\begin{eqnarray}\label{k2}
A_\mu=A_t dt+A_\phi d\phi,
\end{eqnarray}
where
\begin{eqnarray*}
A_t=\frac{Q}{r}, A_{\phi}=-\frac{Q a \sqrt{1+L} \sin^2 \theta}{r}.
\end{eqnarray*}
The expression $\Delta_r$ can be factorised as
\begin{eqnarray*}
\Delta_r=\frac{(1+L)\Lambda}{3}(r_+-r)(r-r_h)(r-r_-)(r-r_{--}).
\end{eqnarray*}
The four apparent singularities of slowly rotating KNdS-like black hole in bumblebee gravity model are obtained only when $\Delta_r=0$. The four horizons are represented by $r_+$, $r_h$, $r_-$ and $r_{--} \,\,(r_+> r_h> r_->r_{--})$. The biggest root $r_+$ indicates the location of cosmological horizon, $r_h$ represents the location of event horizon and $r_-$ corresponds to the location of Cauchy horizon. If one reaches singularity $r=0$, $\theta=\pi/2$, the other side of $r=0, r=r_{--}$ is taken as another cosmological horizon. Fig. \ref{delta} shows the variation of  horizon's position for different values of $L$. The vanishing points represent the position of the horizons. One can clearly see that the distance between the horizon reduces with increasing the Lorentz violation parameter. 
Applying the transformation as $\varphi=\phi-\Omega t$ where the angular velocity $\Omega=-\frac{g_{t\phi}}{g_{\phi\phi}}$, the metric of Eq. \eqref{so} can be transformed to
\begin{eqnarray}\label{k3}
ds^2&=&f(r) dt^2-\frac{1}{g(r)} dr^2-r^2 d\theta^2
- r^2  \sin^2 \theta d\varphi^2,
\end{eqnarray}
where $f(r)=\frac{\Delta_r}{r^2}$ and $g(r)=\frac{\Delta_r}{(1+L)r^2}$.
The surface gravity near the event horizon of slowly rotating KNdS-like black hole is given by
\begin{eqnarray}\label{k4}
&\kappa=&\lim_{f\rightarrow 0} \Big(-\frac{1}{2} \sqrt{-\frac{g}{f}} \frac{d f}{dr} \Big)
= \frac{-M-\frac{r_h \Lambda}{3}(1+L)r_h^2+r_h(1-\frac{1}{3}(1+L)r_h^2) \Lambda}{r_h^2\sqrt{1+L}}.
\end{eqnarray}
The angular velocity near the event horizon of slowly rotating KNdS-like black hole is defined by
\begin{eqnarray}\label{k5}
\Omega=\frac{a}{r_h^2}.
\end{eqnarray}
\section{Fermions tunneling of slowly rotating KNdS-like black hole under GUP}
To study the modified Hawking temperature near the event horizon of slowly rotating KNdS-like black hole under the influence of  GUP, we use the modified Dirac equation with mass $m$, charge $e$ and quantum gravity effect parameter $\beta$, which can be written as \cite{ibungo}
\begin{eqnarray}\label{t1}
[i \gamma^0 \partial_0+i \gamma^i \partial_i(1-\beta m^2)+i\gamma^i \beta \hbar^2(\partial_j \partial^j)\partial_i+\frac{m}{\hbar}(1+\beta \hbar^2 \partial_j \partial^j-\beta m^2)\cr
-\gamma^{\mu} \frac{e}{\hbar} \hat{A_{\mu}}(1+\beta \hbar^2 \partial_j \partial^j-\beta m^2)+i \gamma^{\mu} \Omega_{\mu}(1+\beta \hbar^2 \partial_j \partial^j-\beta m^2)]\Psi=0,
\end{eqnarray}
where $\Omega_{\mu}=\frac{1}{2}i \Gamma_{\mu}^{\nu \lambda}\Sigma_{\nu \lambda}, \,\, \Sigma_{\nu \lambda}=\frac{1}{4}i[\gamma^{\mu},\gamma^{\nu}]$ and $\{\gamma^{\mu},\gamma^{\nu} \}=2g^{\mu \nu}I,$ with $\mu, \nu, \lambda=0, 1, 2$ and $i, j=1,2$. From Eq. \eqref{k3}, the gamma matrices $\gamma^{\mu}$ can be constructed as
\begin{eqnarray}\label{t2}
\gamma^t &=&\frac{1}{\sqrt{f(r)}}
 \begin{bmatrix} 
	i&0&0&0\\
	0&i&0&0\\
	0&0&-i&0\\
	0&0&0&-i\\
	\end{bmatrix},
	\nonumber\\
	\gamma^r &=&\sqrt{g(r)}
 \begin{bmatrix} 
	0&0&1&0\\
	0&0&0&-1\\
	1&0&0&0\\
	0&-1&0&0\\
	\end{bmatrix},
	\nonumber\\
	\gamma^\theta &=&\frac{1}{\sqrt{g_{\theta \theta}}}
 \begin{bmatrix} 
	0&0&0&1\\
	0&0&1&0\\
	0&1&0&0\\
	1&0&0&0\\
	\end{bmatrix},
	\nonumber\\
	\gamma^\varphi &=&\frac{1}{\sqrt{g_{\varphi \varphi}}}
 \begin{bmatrix} 
	0&0&0&-i\\
	0&0&i&0\\
	0&-i&0&0\\
	i&0&0&0\\
	\end{bmatrix}.
	\nonumber\\
	\end{eqnarray}
	The significance of solving Dirac equation is to derive the imaginary part of the radiant action of fermions which, in term, is connected to the Boltzmann factor of emission according to semiclassical WKB approximation.
The wave function for spin up particle can be written as
\begin{eqnarray}\label{t3}
	\psi=\begin{bmatrix}
	A\\ 0 \\B \\ 0
	\end{bmatrix}  \rm exp \frac{i}{\hbar}I(t,r,\theta,\varphi),
	\end{eqnarray}
where $A, B$ and $I$ denote the functions of coordinates $t, r, \theta$ and $\varphi$. $I$ represents the action of emitted fermions. Chandrasekhar \cite{chandra} showed that decoupled of Dirac equation can be done for stationary spacetime or in the spherically symmetric Vaidya-Bonner spacetime \cite{bonner}. Substituting Eqs. \eqref{t2} and \eqref{t3} in Eq. \eqref{t1}, ignoring the higher terms of $\beta$ and $\hbar$, the four decouple equations can be derived as
\begin{eqnarray}\label{t4}
&& -\frac{1}{\sqrt{f}}i A \partial_t I-\sqrt{g}(1-\beta m^2)B \partial_r I+\sqrt{g} B \beta \Big(g(\partial_r I)^2+g^{\theta\theta}(\partial_{\theta}I)^2+g^{\varphi \varphi}(\partial_\varphi I)^2 \Big)\partial_r I \cr
&&+m(1-\beta m^2)A-m \beta \Big(g(\partial_r I)^2+g^{\theta\theta}(\partial_{\theta}I)^2+g^{\varphi \varphi}(\partial_\varphi I)^2 \Big)A-\frac{e}{\sqrt{f}}i A A_t(1-\beta m^2) \cr
&&+\frac{e}{\sqrt{f}}i A A_t\Big(g(\partial_r I)^2+g^{\theta\theta}(\partial_{\theta}I)^2+g^{\varphi \varphi}(\partial_\varphi I)^2 \Big)=0,
\end{eqnarray}
\begin{eqnarray}\label{t5}
&& \frac{1}{\sqrt{f}}i B \partial_t I-\sqrt{g}A(1-\beta m^2) \partial_r I+\sqrt{g} A \beta \Big(g(\partial_r I)^2+g^{\theta\theta}(\partial_{\theta}I)^2+g^{\varphi \varphi}(\partial_\varphi I)^2 \Big)\partial_r I \cr
&&+B m[(1-\beta m^2)
-\beta \Big(g(\partial_r I)^2+g^{\theta\theta}(\partial_{\theta}I)^2+g^{\varphi \varphi}(\partial_\varphi I)^2 \Big)]+\frac{e}{\sqrt{f}}i B A_t(1-\beta m^2) \cr
&& - \frac{e}{\sqrt{f}}i B A_t \Big(g(\partial_r I)^2+g^{\theta\theta}(\partial_{\theta}I)^2+g^{\varphi \varphi}(\partial_\varphi I)^2  \Big)=0,
\end{eqnarray}
\begin{eqnarray}\label{t6}
&& -\frac{1}{\sqrt{g_{\theta\theta}}}B(1-\beta m^2)\partial_{\theta}I-\frac{1}{\sqrt{g_{\varphi \varphi}}}i B(1-\beta m^2)\partial_{\varphi}I+\frac{1}{\sqrt{g_{\theta\theta}}}B \beta \Big(g(\partial_r I)^2+g^{\theta\theta}(\partial_{\theta}I)^2 \cr
&&+g^{\varphi \varphi}(\partial_\varphi I)^2 \Big)\partial_{\theta}I
+\frac{1}{\sqrt{g_{\varphi \varphi}}}i B \beta \Big(g(\partial_r I)^2+g^{\theta\theta}(\partial_{\theta}I)^2+g^{\varphi \varphi}(\partial_\varphi I)^2 \Big)\partial_{\varphi}I-\frac{e}{\sqrt{g_{\varphi \varphi}}}A_\varphi i B \cr
&& \times (1-\beta m^2) +\frac{e}{\sqrt{g_{\varphi \varphi}}}i B A_\varphi\Big(g(\partial_r I)^2+g^{\theta\theta}(\partial_{\theta}I)^2+g^{\varphi \varphi}(\partial_\varphi I)^2 \Big)=0,
\end{eqnarray}
\begin{eqnarray}\label{t7}
&&-\frac{1}{\sqrt{g_{\theta\theta}}}A(1-\beta m^2)\partial_{\theta}I-\frac{1}{\sqrt{g_{\varphi \varphi}}}i A (1-\beta m^2)\partial_{\varphi}I +\frac{1}{\sqrt{g_{\theta\theta}}} A \beta \Big(g(\partial_r I)^2+g^{\theta\theta}(\partial_{\theta}I)^2 \cr
&&+g^{\varphi \varphi}(\partial_\varphi I)^2 \Big)\partial_{\theta}I+\frac{1}{\sqrt{g_{\varphi \varphi}}}i A \beta \Big(g(\partial_r I)^2+g^{\theta\theta}(\partial_{\theta}I)^2+g^{\varphi \varphi}(\partial_\varphi I)^2 \Big)\partial_{\varphi}I \cr
&&-\frac{e}{\sqrt{g_{\varphi \varphi}}}A_\varphi i A(1-\beta m^2)
+\frac{e}{\sqrt{g_{\varphi \varphi}}}i A A_\varphi\Big(g(\partial_r I)^2+g^{\theta\theta}(\partial_{\theta}I)^2+g^{\varphi \varphi}(\partial_\varphi I)^2 \Big)=0.
\end{eqnarray}
To discuss the Hawking radiation of black hole, the variables can be separated as
\begin{eqnarray}\label{t8}
I=-\omega t+\hat{R}_0(r)+Z(\theta,\varphi),
\end{eqnarray}
where $\omega$ denotes the energy of the emitted fermions. Using Eq. \eqref{t8} in Eqs. \eqref{t4}-\eqref{t7}, the two radial equations after some calculations are derived as
\begin{eqnarray}\label{t9}
&& A \Big[\frac{i \omega}{\sqrt{f}}-\frac{e A_t i}{\sqrt{f}}[1-\beta m^2
- \beta g (\partial_r \hat{R}_0)^2]+m(1-\beta m^2)-m \beta g (\partial_r \hat{R}_0)^2 \Big] \cr
&&+B \Big[-\sqrt{g}(1-\beta m^2)(\partial_r \hat{R}_0)+\sqrt{g} \beta g(\partial_r \hat{R}_0)^3 \Big]=0
\end{eqnarray} 
and
\begin{eqnarray}\label{t10}
&& A \Big[-\sqrt{g}(1-\beta m^2)(\partial_r \hat{R}_0)+\sqrt{g} \beta g (\partial_r \hat{R}_0)^3 \Big]+B \Big[-\frac{i \omega}{\sqrt{f}}+m(1-\beta m^2) \cr
&& -m \beta g (\partial_r \hat{R}_0)^2+\frac{e A_t i}{\sqrt{f}}[1-\beta m^2
-\beta g (\partial_r \hat{R}_0)^2] \Big]=0.
\end{eqnarray}
We shall solve the Eqs. \eqref{t9} and \eqref{t10} which determine the modified Hawking temperature of slowly rotating KNdS-like black hole in bumblebee gravity. The non-trivial solution from Eqs. \eqref{t9} and \eqref{t10} can be derived if the determinant of coefficient matrix $A$ and $B$ is zero. That is
\begin{eqnarray}\label{t11}
H_6(\partial_r \hat{R}_0)^6+H_4(\partial_r \hat{R}_0)^4+H_2(\partial_r \hat{R}_0)^2+H_0=0,
\end{eqnarray}
where
\begin{eqnarray}
H_6&=& \beta^2 g^3 g_{tt}, \cr
H_4&=& f g^2 \beta (m^2 \beta-2)-e^2 A_t^2 \beta^2 g^2, \cr
H_2&=& f g(1-\beta m^2)(1+\beta m^2)+2\beta g e A_t \times [-\omega+e A_t(1-\beta m^2)], \cr
H_0&=& -m^2 f(1-\beta m^2)^2-[\omega-e A_t(1-\beta m^2)]^2.
\end{eqnarray}
Neglecting higher power of $\beta$ in Eq. \eqref{t11} and integrating with the help of Feynman prescription and residue theorem of complex analysis near the event horizon $r=r_h$, the two roots having physical meaning are
\begin{eqnarray}\label{t12}
\hat{R}_0^{\pm}&=&\pm \int\sqrt{\frac{m^2 f+[(\omega-j\Omega)-e A_t (1-\beta m^2)]^2}{f g}} \Big(1+\beta \frac{m^2 f+\omega_0^2-e A_t \omega_0}{f}\Big)dr \cr
&=& \pm \frac{i \pi \bar{\omega}_0 r_h^2 \sqrt{1+L}}{\Delta^{'}_r(r_h)}\Big\{1+\beta \Big(m^2+\frac{Y}{r_h^4 \Delta^{'}_r(r_h)}\Big)\Big\},
\end{eqnarray}
where $+/-$ represent to the outgoing/ingoing solutions and
\begin{eqnarray}
Y&=&-2r_h \bar{\omega}_0 r_h^4(\bar{\omega}_0-e A_{t_h})+r_h^2 \Big[r_h^4\{\bar{\omega}_0(-j \Omega_h^{'}-2e A_{t_h}^{'})+(\bar{\omega}_0-e A_{t_h})(-j \Omega_h^{'}-e A_{t_h}^{'})\} \cr&&
+6 \bar{\omega}_0 (\bar{\omega_0}-e A_{t_h})r_h^3 +r_h^4(\bar{\omega}_0-e A_{t_h})\Big \{\frac{m^2 \Delta^{'}_r(r_h)}{2\bar{\omega}_0 r_h^2}+(-j \Omega_h^{'}-2e A_{t_h}^{'}) \Big\} \Big]. \nonumber\\
\end{eqnarray}
Here $A_{t_h}=A_t(r_h),\,\, \Omega_h=\Omega(r_h),\,\, \bar{\omega}_0=\omega-j \Omega_h -e A_{t_h}$ and prime denotes derivative with respect to radial coordinate $r$.
The tunneling rate of fermions which crosses the black hole event horizon in accordance with WKB approximation is derived as
\begin{eqnarray}\label{t13}
\Gamma&=&\rm exp[-2Im \hat{R}_0^{+}+2Im \hat{R}_0^{-}]\cr
&=&\rm exp \Big[-\frac{4\pi \bar{\omega}_0 r_h^2 \sqrt{1+L}}{\Delta^{'}_r(r_h)}\Big\{1+\beta \Big(m^2+\frac{Y}{r_h^4 \Delta^{'}_r(r_h)}\Big)\Big \}\Big].
\end{eqnarray}
The Boltzmann factor gives the Hawking temperature of black hole. The Hawking temperature of slowly rotating KNdS-like black hole in bumblebee gravity under GUP is calculated as 
\begin{eqnarray}\label{t14}
T_h&=&T_0 \Big\{1+\beta \Big(m^2+\frac{Y}{r_h^4 \Delta^{'}_r(r_h)}\Big)\Big \}^{-1}\cr
&=& T_0 \Big\{1-\beta \Big(m^2+\frac{Y}{r_h^4 \Delta^{'}_r(r_h)}\Big)\Big \},
\end{eqnarray}
where 
\begin{eqnarray*}
T_0= \frac{-M-\frac{r_h^3 \Lambda}{3}(1+L)+r_h(1-\frac{\Lambda}{3}(1+L)r_h^2) }{2\pi r_h^2\sqrt{1+L}}.
\end{eqnarray*}
$T_0$ represents the original Hawking temperature of slowly rotating KNdS-like black hole in bumblebee gravity in the absence of quantum gravity effects. Eq. \eqref{t14} indicates that the modified Hawking temperature is smaller than the original one. It shows that a small correction term to the Hawking temperature is produced during the black hole evaporation due to quantum gravity effects. It is noted that the quantum gravity effects prevent the rise of Hawking temperature of black hole which tends to the formation of remnants in black hole physics as shown in \cite{chen1,chen2}. The modified Hawking temperature of slowly rotating KNdS-like black hole in bumblebee gravity depends not only on mass of the black hole, Lorentz violating parameter, electromagnetic potential but also on the mass and energy of the emitted fermions.
\section{Entropy correction of slowly rotating KNdS-like black hole in bumblebee gravity}
To discuss a more accurate results of Hawking radiation temperature and include the quantum correction effects, we need to modify the particle energy correction and the radiation action in terms of different power of $\hbar$ \cite{ran1,ran2,ran3,tan1,tan2,ibungo2,media} as follows
\begin{eqnarray}\label{n1}
\hat{E}=\hat{E}_0+\sum_{i=1}^{\infty} \hbar^i \hat{E}_i
\end{eqnarray}
and
\begin{eqnarray}\label{n2}
\hat{R}=\hat{R}_0+\sum_{i=1}^{\infty} \hbar^i \hat{R}_i,
\end{eqnarray}
where $\hat{E}_0=\omega-\omega_0$ denotes the energy and $\hat{E}_0$, $\hat{R}_0$  are the semi-classical parts of $\hat{E}$ and $\hat{R}$. Here,
\begin{eqnarray}
\hbar^0 && :\frac{d \hat{R}^{\pm}_0}{dr}|_{r\rightarrow r_h}=\pm \frac{\hat{E}_0}{\hat{d}},\label{n3} \\
\hbar^1 && : \frac{d \hat{R}^{\pm}_1}{dr}|_{r\rightarrow r_h}=\pm \frac{\hat{E}_1}{\hat{d}}, \label{n4} \\
\vdots && : \hdots ,
\end{eqnarray}
where 
\begin{eqnarray}\label{n5}
\hat{d}=\frac{\Delta^{'}_r(r_h)}{r_h^2 \sqrt{1+L}} \Big[1+\beta \Big\{m^2+\frac{Y}{r_h^4 \Delta^{'}_r(r_h)}\Big\} \Big]^{-1}.
\end{eqnarray}
 To derive the two dimensional element of the black hole, we take $dt = dr = 0$. Then Eq. \eqref{k3} can be written as
\begin{eqnarray}\label{n6}
ds^2=-r_h^2 d{\theta}^2-r_h^2 \sin^2\theta d\varphi^2.
\end{eqnarray}
The determinant of the covariant metric of Eq. \eqref{n6} is derived as
\begin{eqnarray}\label{n7}
g_{\lambda}=
\begin{vmatrix}
g_{33}&g_{34}\\
g_{43}& g_{44}\\
\end{vmatrix}
=g_{33} g_{44}=r_h^4 \sin^2\theta.
\end{eqnarray}
The black hole area near the event horizon of $r=r_h$ is calculated as
\begin{eqnarray}\label{n8}
A_{h}=\int \sqrt{g_\lambda} \,\, d \theta d \varphi=4 \pi r_h^2.
\end{eqnarray}
It is noted that the black hole entropy is proportional to the horizon area. Thus, the black hole entropy $S_h$ near the event horizon of slowly rotating KNdS-like black hole in bumblebee gravity is calculated as
\begin{eqnarray} \label{n9}
S_{h}=\frac{A_h}{4}=\pi r_h^2.
\end{eqnarray}
 It is known that the solutions of Eqs. \eqref{n3} and \eqref{n4} remain the same, indicating that the solutions depend on each other. This shows that $\hat{R}_i(r)$ is connected to $\hat{R}_0(r)$. Let $\alpha_i$ denote the proportionality constant between $\hat{R}_{i}$ and $\hat{R}_{i+1}$. Then $\hat{R}$ can be expressed as follows
\begin{eqnarray}\label{n10}
\hat{R}=\hat{R}_0+\displaystyle\sum_i \alpha_i \hbar^i \hat{R}_0=\hat{R}_0 \Big[1+\displaystyle\sum_i \frac{\beta_i \hbar^i}{r_h^2} \Big],
\end{eqnarray}
\begin{figure}[!htb]
\centering
\includegraphics[width=175pt,height=155pt]{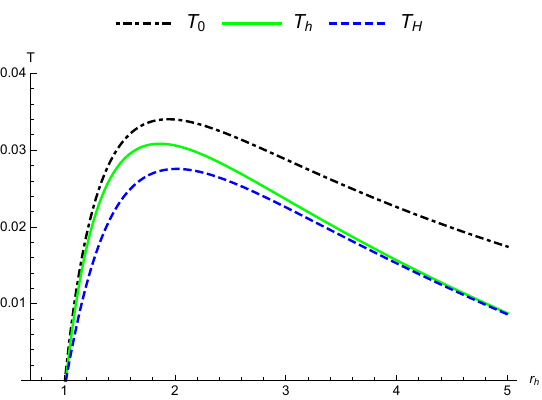}
\caption{Plot of $T_0, T_h$ and $T_H$ with the radius of event horizon $r_h$. Here, $Q=0.1, a=0.1, m=0.2, M=1, \Lambda=0.01, L=0.2, \omega=0.2, j=0.01, e=0.02, \beta=0.1, \beta_i \hbar^i=0.4$. }
\label{HT}
\end{figure}
where $ \beta_i=\alpha_i r_h^2$. To derive an accurate tunneling radiation of fermions, we substitute $\hat{R}_0^{\pm}$ into Eq. \eqref{n3} and the tunneling rate of fermion is derived as
\begin{eqnarray}\label{n11}
\Gamma ' &\sim& \rm{exp}[-2(\rm{Im} \hat{R}_+-\rm{Im} \hat{R}_-)]\cr&
&=\rm{exp}[-\frac{(\omega-\hat{\omega}_0)}{T_{H}}],
\end{eqnarray}
where 
\begin{eqnarray}\label{n12}
T_{H}&=&T_0 \Big[1-\displaystyle \sum_i \frac{\beta_i \hbar^i}{r_h^2} \Big]\Big[1+\beta \Big\{m^2+\frac{Y}{r_h^4 \Delta^{'}_r(r_h)}\Big\} \Big]^{-1} \cr
&=&T_0 \Big[1-\displaystyle \sum_i\frac{\beta_i \hbar^i}{r_h^2} \Big]\Big[1-\beta \Big\{m^2+\frac{Y}{r_h^4 \Delta^{'}_r(r_h)} \Big\} \Big] \cr
&=&T_h \Big[1-\displaystyle \sum_i\frac{\beta_i \hbar^i}{r_h^2} \Big].
\end{eqnarray}
From Eq. \eqref{n12}, $T_H$ indicates the modified Hawking temperature near the event horizon $r=r_h$ of slowly rotating KNdS-like black hole in bumblebee gravity beyond the semiclassical approximation which is connected to the correction of entropy of black hole.
$T_H$ relies on mass, charge, cosmological constant, rotational parameters and Planck constant. We see from Fig. \ref{HT} that the Hawking temperatures $T_H, T_h$ and $T_0$ hold the inequality $T_H<T_h<T_0$ and $T_h$ tends to $T_H$ for large values of event horizon, $r_h$. It is noted that the quantum gravity effects tend to decrease the Hawking temperature of black hole.\\

To investigate the correction of black hole entropy of slowly rotating KNdS-like black hole in bumblebee gravity on the basis of modified Hawking temperature calculated by applying the tunneling of fermion beyond the semiclassical approximation, we shall use the laws of black hole thermodynamics. In accordance with the first law of thermodynamics, the black hole mass M, Hawking temperature $T$, angular momentum $J$, entropy $S$, angular velocity $\Omega$ and electrostatic potential $V$ of slowly rotating KNdS-like black hole in bumblebee gravity hold the condition
\begin{eqnarray}\label{e}
dM=T dS+\Omega dJ+V dQ,
\end{eqnarray}
where $V$ and $\Omega$ are constant near the event horizon of black hole.
Integrating Eq. \eqref{e}, we derive the correction of black hole entropy as
\begin{eqnarray}\label{e11}
S_{bh}&=&\int \frac{dM-\Omega dJ-VdQ}{\hat{T}_{H}}\cr
&=&\int \frac{dM-\omega dJ-VdQ}{T_0} \Big[1+\beta \Big\{m^2+\frac{Y}{r_h^4 \Delta^{'}_r(r_h)}\Big\} \Big] \Big[1+\displaystyle\sum_i \frac{\beta_i \hbar^i}{r_h^2}\Big]\cr
&=&\int dS_{h}\Big[1+\beta \Big \{m^2+\frac{Y}{r_h^4 \Delta^{'}_r(r_h)}\Big \} \Big] \Big[1+\displaystyle\sum_i \frac{\beta_i \hbar^i}{r_h^2}\Big].
\end{eqnarray}
From Eqs. \eqref{n3} and \eqref{n4},
\begin{eqnarray}
\hbar^0 && :\int dS_{h} \Big[1+\beta \Big \{m^2+\frac{Y}{r_h^4 \Delta^{'}_r(r_h)}\Big \} \Big]
=\Big[1+\beta \Big \{m^2+\frac{Y}{r_h^4 \Delta^{'}_r(r_h)}\Big \} \Big]S_{h}\cr
&& ~~ \quad \quad \quad \quad \quad \quad \quad \quad \quad \quad \quad \quad \quad   \quad= \pi r_h^2\Big[1+\beta \Big \{m^2+\frac{Y}{r_h^4 \Delta^{'}_r(r_h)}\Big \} \Big], \label{e1}\\
\hbar^1 &&:\int dS_{h}\frac{\beta_1}{r_h^2} \Big[1+\beta \Big \{m^2+\frac{Y}{r_h^4 \Delta^{'}_r(r_h)}\Big \} \Big] =\pi \beta_1 ln S_{h}\Big[1+\beta \Big \{m^2+\frac{Y}{r_h^4 \Delta^{'}_r(r_h)}\Big \} \Big]\cr
&&   ~~  \quad \quad \quad \quad \quad \quad \quad \quad \quad \quad \quad \quad \quad  \quad =\beta_1^{'} ln S_{h}\Big[1+\beta \Big \{m^2+\frac{Y}{r_h^4 \Delta^{'}_r(r_h)}\Big \} \Big], \label{e2}\cr
 \vdots && : \hdots
\end{eqnarray}
where $\beta_i^{'}=\pi \beta_i$. Continuing in this way, we get
\begin{eqnarray}
\hbar^2 &&:\int dS_{h}\frac{\beta_2}{r_h^2} \sqrt{\frac{1-\lambda c_t^2}{1+\lambda c_r^2}}=\beta_2^{'} ln S_{h}\Big[1+\beta \Big \{m^2+\frac{Y}{r_h^4 \Delta^{'}_r(r_h)}\Big \} \Big], \label{e3} \nonumber \\
\vdots && : \hdots
\end{eqnarray}
Using Eqs. \eqref{e1}, \eqref{e2} and \eqref{e3} in Eq. \eqref{e11}, the accurate modified entropy of the black hole can be obtained as
\begin{eqnarray}
S_{bh}&=&(S_{h}+\beta_1^{'} ln S_{h}+\beta_2^{'} ln S_{h} +...)\Big[1+\beta \Big \{m^2+\frac{Y}{r_h^4 \Delta^{'}_r(r_h)}\Big \} \Big].
\end{eqnarray}
The above equation indicates the correction of entropy beyond the semiclassical approximation for the tunneling of fermions near the event horizon of slowly rotating KNdS-like black hole in bumblebee gravity. The first term on the right hand side of the above equation indicates the original Bekenstein-Hawking entropy while the second and remaining terms are originating from quantum perturbation theory correction. This shows that the entropy of black hole after quantum correction depends on the results of semiclassical theory and Planck constant. 

\section{Scalar perturbation}
To discuss the effective potential and scalar perturbation of slowly rotating KNdS-like black hole in bumblebee gravity, we consider the Klein-G\"{o}rdon equation for massive and charge scalar field perturbation in curve spacetime \cite{li3,ran,ablu2}
\begin{eqnarray}\label{s1}
\frac{1}{\sqrt{-g}}(\partial_a- i q A_a)[\sqrt{-g} g^{ab}(\partial_b- i q A_b)]\Psi=\mu_0^2 \Psi,
\end{eqnarray}
where $\mu_0$ and $q$ are the mass and charge of the scalar field perturbation respectively. Using Eq. \eqref{so} in Eq. \eqref{s1}, we get
\begin{eqnarray}\label{s2}
&&\frac{1}{(1+L)}\partial_r(\Delta_r \partial_r \Psi)-\frac{r^4}{\Delta_r}\partial_t^2 \Psi
+\partial_\theta^2 \Psi+\frac{1}{ \sin^2\theta}\partial_\phi^2 \Psi-\frac{ a r^2 \sqrt{1+L}}{\Delta_r} \partial_t \partial_\phi \Psi +a\sqrt{1+L} \,\, \partial_t \partial_\phi \Psi  \cr&&
+\frac{2i q Q r^3}{\Delta_r}\partial_t \Psi +\frac{2i q Q r a\sqrt{1+L}}{\Delta_r}\partial_\phi \Psi +q^2 Q^2 \Psi-2q^2 Q^2 a \sqrt{1+L} \sin^2\theta \,\, \Psi+\mu_0^2 r^2 \Psi=0 .\nonumber\\
\end{eqnarray}
Since the Hawking radiation takes place along the radial direction only, we use the following ansatz for the separations of the variables $t, r, \theta$ and $\phi$ contained in the above equation as
\begin{eqnarray}\label{s3}
\Psi(r,t)=e^{-i \omega t}R(r) S(\theta, \phi),
\end{eqnarray}
where $\omega$ is the energy of the particle. Then Eq. \eqref{s2} can be written as
\begin{eqnarray}\label{s4}
&&\frac{1}{R(1+L)}\partial_r(\Delta_r \partial_r R)+\frac{\omega^2 r^4}{\Delta_r}+\frac{1}{S} \partial_\theta^2 S+\frac{1}{S \sin^2 \theta}\partial_\phi^2 S+\frac{i \omega r^2 a \sqrt{1+L}}{S \Delta_r}\partial_\phi S-\frac{i \omega a \sqrt{1+L}}{S} \partial_\phi S \cr&&
+\frac{2\omega q Q r^3}{ \Delta_r}+\frac{2i q Q r a \sqrt{1+L}}{S \Delta_r}\partial_\phi S
+q^2 Q^2-2q^2 Q^2 a \sqrt{1+L} \sin^2\theta
+\mu_0^2 r^2=0.
\end{eqnarray}
Therefore, the radial part from Eq. \eqref{s4} is derived as
\begin{eqnarray}\label{s5}
&&\frac{1}{R(1+L)}\frac{\partial}{\partial r}\Big(\Delta_r \frac{\partial R}{\partial r}\Big)+\frac{\omega^2 r^4}{\Delta_r}+\frac{2 r^3 \omega q Q}{\Delta_r}+q^2 Q^2+\mu_0^2 r^2-\lambda=0,
\end{eqnarray}
where $\lambda=\ell(\ell+1),\,\, \ell$ is the angular harmonic index.
To derive a one-dimensional wave equation from the above equation, we use the tortoise coordinate $r_*$, defined by
\begin{eqnarray}\label{s6}
\frac{dr_*}{dr}=\frac{r^2\sqrt{1+L}}{\Delta_r},
\end{eqnarray}
together with the following transformation 
\begin{eqnarray}\label{s7}
R(r)=\frac{U(r)}{r}.
\end{eqnarray}
Using Eqs. \eqref{s6} and \eqref{s7} in Eq. \eqref{s5}, we obtain a one-dimensional Schr{\"o}dinger like wave equation as
\begin{eqnarray}\label{s8}
\frac{d^2U}{dr_*^2}+(\omega^2-V_{eff})U=0,
\end{eqnarray}
where the effective potential $V_{eff}$ of massive scalar perturbation is given by
\begin{eqnarray}\label{s9}
V_{eff}&=&\frac{\Delta_r}{(1+L)r^4}\Big[\frac{\Delta_r^{'}}{r}-\frac{2\Delta_r}{r^2}-\frac{2 \omega q Q r^3 (1+L)}{\Delta_r}-q^2 Q^2 (1+L)-\mu_0^2 r^2(1+L)+\lambda (1+L) \Big].\nonumber\\
\end{eqnarray}
\begin{figure}[!htbp]
\centering
\subfloat[ $Q=0.1, \Lambda=0.01, \ell=1, M=1,\mu_0=0.1,\omega=0.2, q=0.001$.]
{\includegraphics[width=175pt,height=155pt]{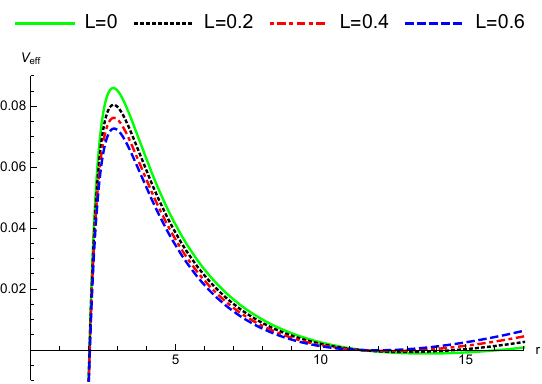}
\label {V1}
}
\hfill
\subfloat[ $Q=0.1, L=0.2, \ell=1, M=1,\mu_0=0.1,\omega=0.2, q=0.001$.]
{\includegraphics[width=175pt,height=155pt]{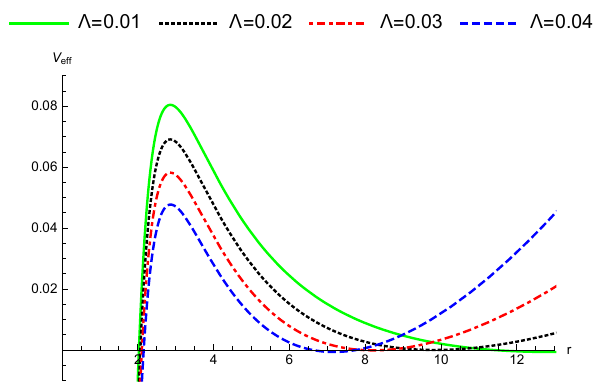}
\label {V2}
}
\hfill
\subfloat[ $Q=0.1, \Lambda=0.01, L=0.2, M=1,\mu_0=0.1,\omega=0.2, q=0.001$.]
{\includegraphics[width=175pt,height=155pt]{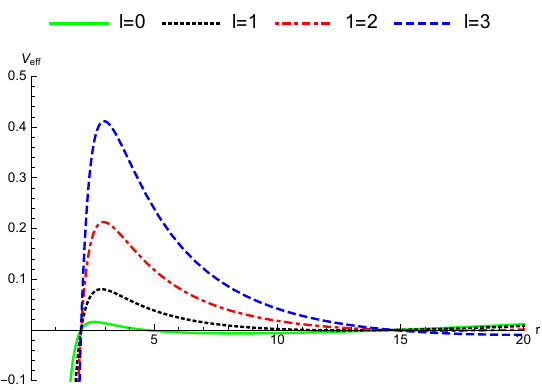}
\label {V3}
}
\hfill
\subfloat[ $ \Lambda=0.01, L=0.2, \ell=1, M=1,\mu_0=0.1,\omega=0.2, q=0.001$.]
{\includegraphics[width=175pt,height=155pt]{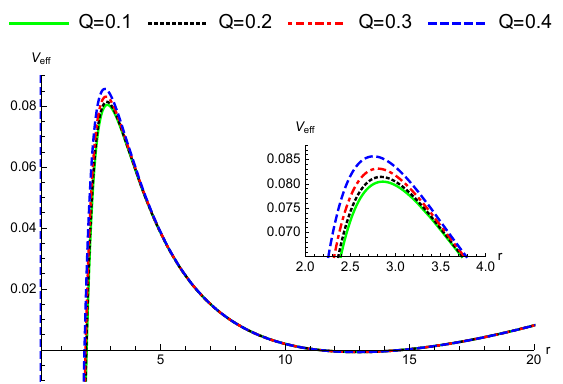}
\label {V4}
}
\caption{The figure plot the effective potential $V_{eff}$ with radial coordinate $r$ for the scalar perturbation varying with different parameters.}
\label{veff}
\end{figure}
From the above equation, we find that the $V_{eff}$ is dependent on the black hole parameters: $Q, M, L, \Lambda, \mu_0$ and $\ell$. If $Q=0$ and $\Lambda=0$, the effective potential of massive slowly rotating KNdS-like black hole in bumblebee gravity reduces to the effective potential of slowly rotating Kerr-like black hole \cite{kanzi2}. Since the QNMs are determined by the effective potential, we first address how the effective potential depends on these parameters. Fig. \ref{veff}  shows the fluctuation of the effective potential of massive scalar particle for different values of $L, \Lambda, \ell$ and $Q$. From Fig. \ref{V1}, it is observed that the effective potential of massive scalar field vanishes at three points: $r_1$, $r_2$ and $r_3$ ($r_1<r_2<r_3$). In the region $r_1<r<r_2$, the effective potential is positive and has a local maximum. The local maximum of the effective potential reduces with increasing the magnitude of $L$. Moreover, the width of the potential decreases as $L$ increases. Since the potential barrier reduces with increasing the values of $L$, the waves become more easier to transmit to the surrounding, indicating that the greybody factors will increase. The details of this claim will be thoroughly examined in the following section. Fig. \ref{V2} shows that for a fixed $L$, the peak of the effective potential reduces with increasing the $\Lambda$ and thus the potential barrier decreases, enhancing the emission of scalar particles. Figs. \ref{V3} and \ref{V4}, we display the dependence of the effective potentials on $\ell$ and $Q$ respectively. The peak of the effective potential are found to increase with increasing $\ell$ and $Q$. However, the position of the peak shifts from left to right with increasing $\ell$ and the situation is reversed with increasing $Q$.
\subsection{Greybody radiation of scalar particle in slowly rotating KNdS-like black hole in bumblebee gravity}
In this section, we study the greybody factor of scalar particles by using the effective potential of slowly rotating KNdS-like black hole in bumblebee gravity. Taking quantum gravity effect into consideration, a black hole radiates the thermal radiation like a black body spectrum which is known as Hawking radiation. While thermal radiation propagates out from the black hole event horizon, it encounters the spacetime curvature of the black hole and this interaction alters the intensity and spectrum of the Hawking radiation which escapes to infinity. Therefore, an observers at infinite distance observes the modified form of thermal radiation which is distinct from the original thermal radiation near the event horizon of black hole. The greybody factor is the amount of wave or particle which tunnel through the potential barrier. The spacetime curvature of a black hole behaves as a potential barrier from a black hole scattering point of view. Hence the greybody factor is also known as transmission coefficient. The rigorous bound on the greybody factor of scalar particle is given by \cite{visser,boon1}
\begin{eqnarray}\label{g1}
\sigma_s(\omega)\geq \sec h^2 \Big(\int_{-\infty}^{+\infty}\wp \,\, dr_* \Big),
\end{eqnarray}
where
\begin{eqnarray*}
\wp=\frac{\sqrt{(h')^2+(\omega^2-V_{eff}-h^2)^2}}{2h}.
\end{eqnarray*}
Here, $h(r_*)$ represents a positive function and it will satisfy the conditions $h(-\infty)=h(\infty)=\omega$. For simplicity, we take $h=\omega$. Thus Eq. \eqref{g1} reduces to
\begin{eqnarray}\label{g2}
\sigma_s(\omega)\geq \sec h^2 \Big(\int_{-\infty}^{+\infty}\frac{V_{eff}}{2\omega} dr_* \Big).
\end{eqnarray}
The tortoise coordinate $r_*$ can be expressed as
\begin{eqnarray}\label{g3}
r_*=\frac{3}{\Lambda \sqrt{1+L}}\Big(\frac{r_+^2 \,\, log[r - r_+]}{(r_+ - r_h) (r_+ - r_-) (r_+ - r_{--})}+\frac{r_h^2 \,\, log[r - r_h]}{(r_+ - r_h) (r_h - r_-) (r_h - r_{--})} \cr
+\frac{r_-^2 \,\, log[r - r_-]}{(r_+ - r_-) (-r_h + r_-) (r_- - r_{--})}+\frac{ r_{--}^2 \,\,\, log[r - r_{--}]}{(r_+ - r_{--}) (-r_h + r_{--}) (-r_- + r_{--})} \Big).
\end{eqnarray}
Let us examine the effective potential in its massless form in order to have an immaculately integration. Here, it can be noted that the tortoise coordinate $r_* \rightarrow -\infty$ and $r_* \rightarrow +\infty$ when $r \rightarrow r_h$ and $r \rightarrow r_+$. Thus, Eq. \eqref{g2} can be written as 
\begin{eqnarray}\label{g4}
\sigma_s(\omega)\geq \sec h^2 \Big(\int_{r_h}^{r_+}\frac{V_{eff}}{2\omega} dr \Big).
\end{eqnarray}

\begin{figure}[!htbp]
\centering
\subfloat[$Q=0.1, \Lambda=0.01, \ell=1, q=0.01, M=1$.]
{\includegraphics[width=175pt,height=155pt]{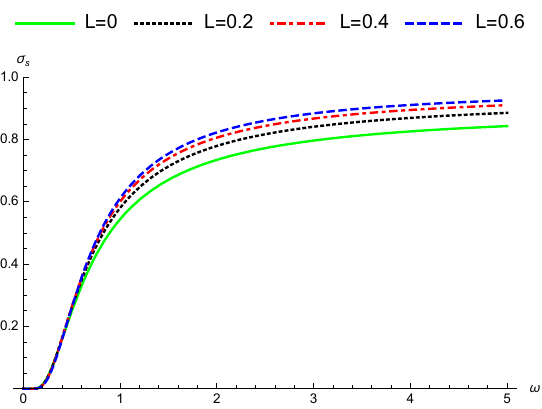}
\label {G1}
}
\hfill
\subfloat[$Q=0.1, L=0.2, \ell=1, q=0.001, M=1.$]
{\includegraphics[width=175pt,height=155pt]{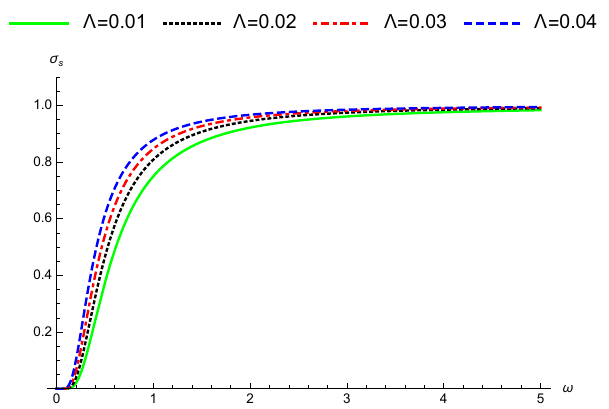}
\label {G2}
}
\hfill
\subfloat[$Q=0.1, \Lambda=0.01, L=0.2, q=0.001, M=1$.]
{\includegraphics[width=175pt,height=155pt]{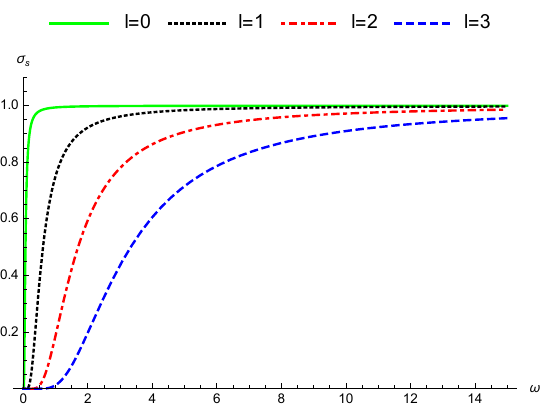}
\label {G3}
}
\hfill
\subfloat[$L=0.2, \Lambda=0.01, \ell=1, q=0.001, M=1.$]
{\includegraphics[width=175pt,height=155pt]{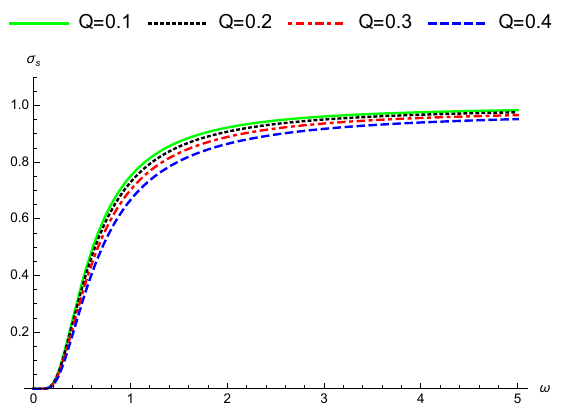}
\label {G4}
}
\caption{Plots of lower bound of the greybody factor $\sigma_s$ with $\omega$ varying with different parameters.}
\end{figure}

Using Eq. \eqref{s9} in Eq. \eqref{g4} and applying rigorous bound on the greybody factor of scalar particle, we get the greybody factor of slowly rotating KNdS-like black hole as
\begin{eqnarray}\label{g5}
\sigma_s(\omega)&\geq& \sec h^2 \Big(\frac{1}{2\omega \sqrt{1+L}}\Big) \int_{r_h}^{r_+}\Big\{\frac{\Delta_r^{'}}{r^3}-\frac{2\Delta_r}{r^4}-\frac{2 \omega q Q r (1+L)}{\Delta_r}-\frac{q^2 Q^2 (1+L)}{r^2}-\mu_0^2 (1+L) \cr&&
+\frac{\lambda (1+L)}{r^2} \Big\}dr.
\end{eqnarray}
Therefore the rigorous bound of the slowly rotating KNdS-like black hole for massless scalar field is calculated as
\begin{eqnarray}\label{g6}
\sigma_s(\omega)&\geq& \sec h^2 \Big(\frac{1}{2\omega \sqrt{1+L}}\Big) \Big[-\frac{2(1+L)\Lambda}{3}(r_+-r_h)+\Big\{\frac{q^2 Q^2(2+3L+L^2)}{2+L}-(1+L)\lambda \Big\}   \cr&& \times \Big(\frac{1}{r_+}-\frac{1}{r_h}\Big) -\frac{3\omega q Q+M \Lambda}{\Lambda} \Big(\frac{1}{r_+^2}-\frac{1}{r_h^2}\Big)+\frac{4Q(1+L)}{3(2+L)}\Big(\frac{1}{r_+^3}-\frac{1}{r_h^3}\Big)-\frac{9\omega q Q}{2(1+L) \Lambda^2} \cr&& \times \Big(\frac{1}{r_+^4}-\frac{1}{r_h^4}\Big) +\frac{36M \omega q Q}{5(1+L) \Lambda^2} \Big(\frac{1}{r_+^5}-\frac{1}{r_h^5}\Big)-\frac{6\omega q Q^3}{(2+L)\Lambda^2} \Big(\frac{1}{r_+^6}-\frac{1}{r_h^6}\Big)
\Big],
\end{eqnarray}
The nature of the greybody factor depends on the effective potential.  The higher of the effective potential lowers the transmission amplitude and it decreases the bound for the greybody factor.  Now let us study this behaviour qualitatively for slowly rotating KNdS-like black hole in bumblebee gravity model. We study the dependence of greybody factors on Lorentz violation parameters by plotting $\sigma_s(\omega)$  numerically for various values of $L$ in Fig. \ref{G1}. It is observed that the rigorous bound of the greybody factor increases with increasing $L$. This indicates that the value of the greybody factor bound increases as the strength  of the effective potential barrier decreases. Moreover, we also investigate the effect of  $\Lambda$, $\ell$ and $Q$ on greybody factor and display in Figs. \ref{G2} to \ref{G4}. The rigorous bound of the greybody factor decreases with increasing $\ell$ and $Q$ but it has an opposite effect with increasing $\Lambda$. It is worth mentioning that the nature of the greybody factors are again in accordance with the behavior of the effective potential.

\section{QNMs of slowly rotating KNdS-like black hole in bumblebee gravity}
In this section, the scalar field perturbation of slowly rotating KNdS-like black hole is used to investigate the QNMs. It is the solution of wave equation Eq. \eqref{s8} which holds specific boundary condition near the horizon of black hole and very far from the black hole. The QNMs satisfy the conditions of  purely ingoing wave near the event horizon and purely outgoing wave near the cosmological horizon or spatial infinity.
\subsection{WKB method}
In this section, the QNMs of slowly rotating KNdS-like black hole in bumblebee gravity will be investigated by using WKB approximation of third-order for the scalar  perturbation. The WKB third-order formula for finding the quasinormal frequency of slowly rotating KNdS-like black hole in bumblebee gravity model with effective potential $V_{eff}(r)$ is defined by
\begin{eqnarray}\label{q1}
\omega^2=\Big[V_0+(-2V_0^{''})^{\frac{1}{2} } \Lambda_*(n)-i(n+\frac{1}{2})(-2V_0^{''})[1+\Omega_*(n)]\Big],
\end{eqnarray}
where
\begin{eqnarray}\label{q2}
\Lambda_*(n)=\frac{1}{(-2V_0^{''})^{\frac{1}{2} }}\Big[\frac{1}{8}\Big(\frac{V_0^{(4)}}{V_0^{''}}\Big)\Big(\frac{1}{4}+\alpha^2 \Big)-\frac{1}{288} \Big(\frac{V_0^{'''}}{V_0^{''}} \Big)^2 (7+60 \alpha^2) \Big]
\end{eqnarray}
and
\begin{eqnarray}\label{q3}
&\Omega_*=\frac{1}{(-2V_0^{''})}\Big[\frac{5}{6912}\Big(\frac{V_0^{'''}}{V_0^{''}} \Big)^4 (77+188\alpha^2)-\frac{1}{384}\Big(\frac{V_0^{'''^2} V_0^{(4)}}{V_0^{''^3}}\Big)(51+100\alpha^2)+\frac{1}{2304}\cr
&\times \Big(\frac{V_0^{(4)}}{V_0^{''}}\Big)^2 (67+68\alpha^2)+\frac{1}{288}\Big(\frac{V_0^{'''} V_0^{(5)}}{V_0^{''^2}} \Big)\times (19+28\alpha^2)-\frac{1}{288}\Big(\frac{V_0^{(6)}}{V_0^{''}}\Big)
(5+4\alpha^2) \Big], \nonumber\\
\end{eqnarray}
where superscripts $(n=4,5,6)$ and primes are the differentiation of the effective potential with respect to the tortoise coordinate $r_*$ and $V_0$ is the effective potential calculated at $r_0$, which indicates the location of the peak of the effective potential and $\alpha=n+\frac{1}{2}$, where $n$ is a positive integer which represents the overtone number. The accuracy of the WKB method are determined by the overtone number $(n)$ and the multipole number $(\ell)$. The third-order WKB method is more accurate and reliable if $\ell>n$ but it is inapplicable when $\ell<n$. Hence $\ell$ and $n$ are critical factor which determines WKB method accurately. Using the third-order WKB approximation, the quasinormal frequencies for the scalar perturbation is obtained for slowly rotating KNdS-like black hole in bumblebee gravity. The results are displayed in Tables 1, 2 and 3 by varying different parameters $L$, $\Lambda$ and  $Q$.

\begin{table}[tbp]
\centering
\caption{The table shows the quasinormal frequencies of third-order WKB approximation and P{\"o}schl-Teller approximation for different values of $L$ with fixed $Q=0.1$, $\Lambda=0.01$, $q=0.001$, $\ell=1$ and $M=1$.}

\begin{tabular}{p{0.5cm} p{0.5cm} p{0.5cm} p{2.8cm} p{2.8cm} p{1.5cm} p{2cm}}
\hline
         $\ell$          &        n           &       $L$            &       WKB third-order           &                  P{\"o}schl-Teller fitting & Error(Re)& Error(-Im)  \\ \hline
\multirow{4}{*}{}  & \multirow{4}{*}{} & \multirow{4}{*}{} & \multirow{4}{*}{} & \multirow{4}{*}{} & \multirow{4}{*}{} \\
  0 & 0 & 0   & 0.0988356-0.1139710$i$ & 0.0977037-0.1133972$i$  & 0.0011319 & 0.0005738 \\
    &   & 0.2 & 0.0890512-0.1037959$i$ & 0.0860315-0.1031375$i$  &0.0030197& 0.0006584   \\
    &   & 0.4 & 0.0813274-0.0958497$i$ & 0.0767112-0.0951003$i$ & 0.0046161 & 0.0007494 \\
    &   & 0.6 & 0.0749997-0.0894065$i$ & 0.0689932-0.0885644$i$ & 0.0060065 & 0.0008421 \\ \hline
\multirow{8}{*}{}  & \multirow{4}{*}{} & \multirow{4}{*}{} & \multirow{4}{*}{} & \multirow{4}{*}{} \\
  1 & 0 & 0   & 0.2758090-0.0949628$i$ & 0.2815305-0.0972507$i$  & 0.0057215 & 0.0022879 \\
    &   & 0.2 & 0.2702428-0.0856113$i$ & 0.2748473-0.0874040$i$  &0.0046045& 0.0017927   \\
    &   & 0.4 & 0.2654589-0.0783121$i$ & 0.2692710-0.0797529$i$ & 0.0038120 & 0.0014408 \\
    &   & 0.6 & 0.2611684-0.0723916$i$ & 0.2643919-0.0735734$i$ & 0.0032235 & 0.0011818 \\ \cline{2-7} 
                   & \multirow{4}{*}{} & \multirow{4}{*}{} & \multirow{4}{*}{} & \multirow{4}{*}{} \\
    & 1 & 0   & 0.2524270-0.2940053$i$ & 0.2815305-0.2917520$i$  & 0.0291035 & 0.0022533  \\
    &   & 0.2 & 0.2505683-0.2637701$i$ & 0.2748473-0.2622121$i$  &0.024279& 0.0015580   \\
    &   & 0.4 & 0.2486260-0.2403768$i$ & 0.2692710-0.2392586$i$ & 0.020645 & 0.0011181 \\
    &   & 0.6 & 0.2465719-0.2215459$i$ & 0.2643919-0.2207202$i$ & 0.0178199 & 0.0008257 \\ \hline

\end{tabular}
\end{table}

\begin{table}[tbp]
\centering
\caption{The table shows the quasinormal frequencies of third-order WKB approximation and P{\"o}schl-Teller approximation for different values of $\Lambda$ with fixed  $Q=0.1$, $L=0.2$, $q=0.001$, $\ell=1$ and $M=1$.}
\begin{tabular}{p{0.5cm} p{0.5cm} p{0.5cm} p{2.8cm} p{2.8cm} p{1.5cm} p{2cm}}
\hline
         $\ell$          &        n           &       $\Lambda$            &       WKB third-order           &                  P{\"o}schl-Teller fitting  & Error(Re) & Error(-Im) \\ \hline
\multirow{4}{*}{}  & \multirow{4}{*}{} & \multirow{4}{*}{} & \multirow{4}{*}{} & \multirow{4}{*}{} \\
 0 & 0 & 0.01 & 0.0890512-0.1037959$i$ & 0.0860315-0.1031375$i$  &0.0030197 & 0.0006584 \\
   &   & 0.02 & 0.0812907-0.1017033$i$ & 0.0667496-0.1000157$i$  &0.0145411& 0.0016876  \\
   &   & 0.03 & 0.0725035-0.0982893$i$ & 0.0459996-0.0955772$i$ & 0.0265039  & 0.0027121 \\
   &   & 0.04 & 0.0628497-0.0933170$i$ & 0.0185838-0.0898043$i$ & 0.0442659 & 0.0035127  \\ \hline
\multirow{8}{*}{}  & \multirow{4}{*}{} & \multirow{4}{*}{} & \multirow{4}{*}{} & \multirow{4}{*}{} \\
  1 & 0 & 0.01 & 0.2702428-0.0856113$i$ & 0.2748473-0.0874040$i$  & 0.0046045 & 0.0017927 \\
    &   & 0.02 & 0.2508305-0.0813200$i$ & 0.2542582-0.0827285$i$  &0.0034277& 0.0014085  \\
    &   & 0.03 & 0.2304003-0.0762715$i$ & 0.2329538-0.0773164$i$ & 0.0025535 & 0.001045 \\
    &   & 0.04 & 0.2087631-0.0703711$i$ & 0.2106794-0.0711087$i$ & 0.0019163& 0.0007377 \\ \cline{2-7} 
                   & \multirow{4}{*}{} & \multirow{4}{*}{} & \multirow{4}{*}{} & \multirow{4}{*}{} \\
    & 1 & 0.01 & 0.2505683-0.2637701$i$ & 0.2748473-0.2622121$i$  & 0.024279 & 0.0015580 \\
    &   & 0.02 & 0.2359010-0.2486460$i$ & 0.2542582-0.2481855$i$  &0.0183573& 0.0004605  \\
    &   & 0.03 & 0.2193342-0.2321460$i$ & 0.2329538-0.2319493$i$ & 0.0136196 & 0.0001967 \\
    &   & 0.04 & 0.2006992-0.2136211$i$ & 0.2106794-0.2133262$i$ & 0.0099802 & 0.0002949 \\ \hline

\end{tabular}
\end{table}

\begin{table}[tbp]
\centering
\caption{The table shows the quasinormal frequencies of third-order WKB approximation and P{\"o}schl-Teller approximation for different values of $Q$ with fixed  $L=0.2$, $\Lambda=0.01$, $q=0.001$, $\ell=1$ and $M=1$.}
\begin{tabular}{p{0.5cm} p{0.5cm} p{0.5cm} p{2.8cm} p{2.8cm} p{1.5cm} p{2cm}}
\hline
         $\ell$          &        n           &       $Q$            &       WKB third-order           &                  P{\"o}schl-Teller fitting &Error(Re) & Error(-Im) \\ \hline
\multirow{4}{*}{}  & \multirow{4}{*}{} & \multirow{4}{*}{} & \multirow{4}{*}{} & \multirow{4}{*}{} \\
 0 & 0 & 0.1 & 0.0890512-0.1037959$i$ & 0.0860315-0.1031375$i$  & 0.0030197 & 0.0006584 \\
   &   & 0.2 & 0.0896530-0.1038977$i$ & 0.0866548-0.1032728$i$  &0.0029982& 0.0006249  \\
   &   & 0.3 & 0.0907158-0.1040215$i$ & 0.0878189-0.1034540$i$ & 0.0028970 & 0.0005675 \\
   &   & 0.4 & 0.0922908-0.1041293$i$ & 0.0895800-0.1036482$i$ & 0.0027108 & 0.0004811 \\ \hline
\multirow{8}{*}{}  & \multirow{4}{*}{} & \multirow{4}{*}{} & \multirow{4}{*}{} & \multirow{4}{*}{} \\
  1 & 0 & 0.1 & 0.2702428-0.0856113$i$ & 0.2748473-0.0874040$i$  & 0.0046045 & 0.0017927 \\
    &   & 0.2 & 0.2719694-0.0857985$i$ & 0.2765568-0.0875834$i$  &0.0045874& 0.0017849   \\
    &   & 0.3 & 0.2749563-0.0860992$i$ & 0.2795155-0.0878707$i$ & 0.0045592 & 0.0017715 \\
    &   & 0.4 & 0.2793602-0.0864977$i$ & 0.2838772-0.0882498$i$ & 0.0045170 & 0.0017520 \\ \cline{2-7} 
                   & \multirow{4}{*}{} & \multirow{4}{*}{} & \multirow{4}{*}{} & \multirow{4}{*}{} \\
    & 1 & 0.1 & 0.2505683-0.2637701$i$ & 0.2748473-0.2622121$i$  & 0.0242790 & 0.0015580 \\
    &   & 0.2 & 0.2523867-0.2642987$i$ & 0.2765568-0.2627502$i$  &0.0241701& 0.0015485   \\
    &   & 0.3 & 0.2555346-0.2651447$i$ & 0.2795155-0.2636122$i$ & 0.0239810 & 0.0015325 \\
    &   & 0.4 & 0.2601884-0.2662536$i$ & 0.2838772-0.2647494$i$ & 0.0236887 & 0.0015042 \\ \hline

\end{tabular}
\end{table}

\subsection{P{\"o}schl-Teller fitting method}
In this section, we will evaluate the QNMs of slowly rotating KNdS-like black hole in bumblebee gravity by using P{\"o}schl-Teller potential approximation \cite{pos,ferr}
\begin{eqnarray}\label{pt}
V_{PT}=\frac{V_0}{\cosh^2 \alpha(r_* - r_{*_{0}})}.
\end{eqnarray}
Here, $V_0=V(r_{*_{0}})$ represents the height and $\alpha>0$ denotes the curvature of the effective potential at the maximum ($r_*=r_{*_{0}}$) which is given by
\begin{eqnarray*}
\alpha^2=-\frac{1}{2V_0} \Big[\frac{d^2 V}{dr_*} \Big]_{r_* \rightarrow r_{*_{0}}}.
\end{eqnarray*}
 Thus, the bound states of $V_{PT}$ is defined by
\begin{eqnarray}\label{pt2}
\Omega_n(V_0,\alpha)=\alpha \Big[-(n+\frac{1}{2})+\Big[\frac{1}{4}+\frac{V_0}{\alpha^2}\Big]^\frac{1}{2} \Big].
\end{eqnarray}
The QNMs $\omega$ after applying inverse transformation is derived as
\begin{eqnarray}\label{pt3}
\omega=\pm \sqrt{V_0-\frac{\alpha^2}{4}}-i \alpha(n+\frac{1}{2}), \,\,\,\,\, \rm (n=0,1,2,3,...).
\end{eqnarray}
From the above equation, we observe that the imaginary part of the QNMs depends on the mode number $n$. The calculated values for positive $n$ are displayed in Tables 1, 2 and 3 for different values of $L$, $\Lambda$ and $Q$ of black hole respectively. It is noted that the calculated values by these two methods have a good agreement with each other. Increasing the values of $L$ and $\Lambda$, both the real part and the magnitude of imaginary part of the complex frequencies decrease with the increasing $n$ in these two methods as shown in Tables 1 and 2. It is also shown in Table 3 that the real part and the magnitude of imaginary part of the complex frequencies evaluated by these two approximations, increase with the increasing $n$ and $Q$.\\

\subsection{Hawking sparsity}
\begin{figure}[!htbp]
\centering
\subfloat[$Q=0.1, \Lambda=0.01, l=1, q=0.001, M=1$.]
{\includegraphics[width=175pt,height=155pt]{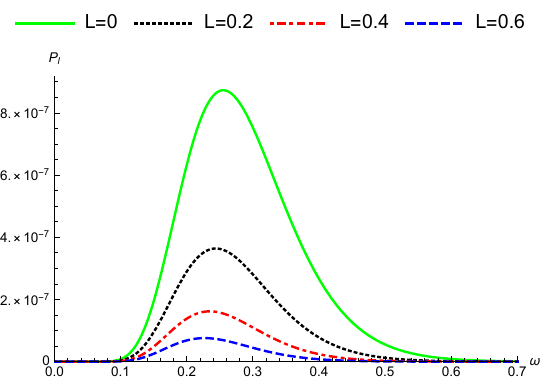}
\label {P1}
}
\hfill
\subfloat[$L=0.2, \Lambda=0.01, l=1, q=0.01, M=1$.]
{\includegraphics[width=175pt,height=155pt]{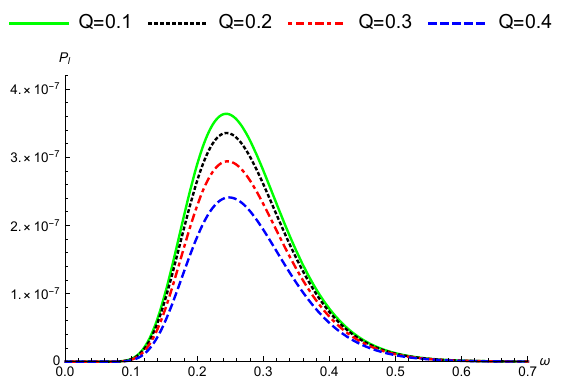}
\label {P2}
}
\hfill
\subfloat[$Q=0.1, L=0.2, l=1, q=0.001, M=1$.]
{\includegraphics[width=175pt,height=155pt]{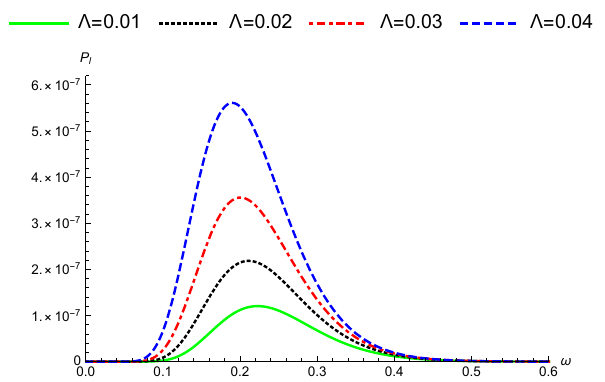}
\label {P3}
}
\caption{Power spectrum of scalar perturbation $P_l$ with $\omega$ for different values of parameters.}
\label{fig:5}
\end{figure}

The effects of Lorentz violating parameter $L$, cosmological constant $\Lambda$ and charge $Q$ on the spectrum and sparsity of Hawking radiation near the event horizon of black hole will be investigated for the tunneling of massless scalar particle. The total power emitted by a black hole in the form of Hawking radiation is expressed as \cite{gray,miao}
\begin{eqnarray}\label{h1}
\frac{d E(\omega)}{dt}\equiv P_{tot}=\sum_{\ell} \sigma_s(\omega) \frac{\omega}{e^{\omega/T_0}-1} \hat{k}. \hat{n} \frac{d^3k\,\, d A}{(2\pi)^3},
\end{eqnarray}
where $\omega$ is the frequency in the momentum interval $d^3k$. \,\,\,$\sigma_s,\,\, \hat{n}$ and $d A$ are the greybody factor, unit normal to $d A$ and the surface element respectively. Taking $|k|=\omega$ for massless particle, the above equation can be written as
\begin{eqnarray}\label{h2}
P_{tot}=\sum_{\ell} \int_0^\infty P_{\ell}(\omega) d\omega.
\end{eqnarray}
Here $P_{\ell}(\omega)$ represents the power emitted per unit frequency in the $\ell^{th}$ mode which is defined by
\begin{eqnarray}\label{h3}
P_{\ell}(\omega)=\frac{A_h}{8\pi^2} \sigma_s(\omega)\frac{\omega^3}{e^{\omega/T_0}-1},
\end{eqnarray}
where $A_h$ denotes the horizon area of slowly rotating KNdS-like black hole. To study the impact of the parameters $L$, $Q$ and $\Lambda$ on the qualitative nature of the power spectrum, we plot $P_\ell$ with respect to $\omega$ in Figs. \ref{P1} to \ref{P3}. The peak of the power spectrum decreases as $L$ and $Q$ increase. However, the peak of the power spectrum increases with increasing $\Lambda$ as shown in Fig. \ref{P3}.
To investigate the thermal radiation emitted by black hole, a dimensionless parameter $\eta$ which is known as the sparsity of Hawking radiation defined as \cite{gray,miao,hod1,hod2}

\begin{table}[htp]
\caption{Numerical values of $\omega_{max}, P_{max}, P_{tot}$ and $\eta$ for scalar perturbation for various values of $L$ with fixed  $Q=0.1$, $\Lambda=0.01$, $q=0.001$, $\ell=1$ and $M=1$.}
\begin{tabular}{p{1.2 cm} p{2.8cm} p{3cm} p{2.9cm} p{2.2cm}}
\toprule
L & \,\,\, $\omega_{max} $ &  \,\,\,\,\,\,$P_{max}$  & \,\,\,\,\,\,\,$P_{tot}$ & \,\,\,\,\,\,\,\,\,\,$\eta$ \\ \midrule
0 & 0.2552416 &  $ 8.7291577\times 10^{-7} $ & $1.7169245 \times 10^{-7}$ &$60390.95 $   \\
0.2 & 0.2439298 & $3.6389473 \times 10^{-7}$ & $ 6.6149201 \times 10^{-8}$ & $143161.18$ \\
0.4 & 0.2347054 &  $1.6168627 \times 10^{-7}$ & $2.7470771 \times 10^{-8}$& $319150.33$  \\
0.6 & 0.2268640  & $7.5619944 \times 10^{-8}$ & $1.2103030 \times 10^{-8}$ & $676795.21 $  \\ \bottomrule
\end{tabular}
\end{table}

\begin{table}[htp]
\caption{Numerical values of $\omega_{max}, P_{max}, P_{tot}$ and $\eta$ for scalar perturbation for various values of $Q$ with fixed  $L=0.2$, $\Lambda=0.01$, $q=0.001$, $\ell=1$ and $M=1$.}
\begin{tabular}{p{1.2 cm} p{2.8cm} p{3cm} p{2.9cm} p{2.2cm}}
\toprule
Q & \,\,\, $\omega_{max} $ &  \,\,\,\,\,\,$P_{max}$  & \,\,\,\,\,\,\,$P_{tot}$ & \,\,\,\,\,\,\,\,\,\,$\eta$ \\ \midrule
0.1 & 0.2439298 &  $ 3.6389473\times 10^{-7}$ & $6.6149201\times 10^{-8}$ &$143161.18 $  \\
0.2 & 0.2445729 & $3.3571112\times 10^{-7}$ & $6.1179318\times 10^{-8}  $ & $155608.13$ \\
0.3 & 0.2460443 &  $2.9410925\times 10^{-7}$ & $5.3816627\times 10^{-8}$ & $179031.87 $  \\
0.4 & 0.2483451  & $2.4107159 \times 10^{-7}$ & $4.4344605\times 10^{-8}$ &$221355.63 $  \\ \bottomrule
\end{tabular}
\end{table}

\begin{table}[htp]
\caption{Numerical values of $\omega_{max}, P_{max}, P_{tot}$ and $\eta$ for scalar perturbation for various values of $\Lambda$ with fixed  $Q=0.1$, $L=0.2$, $q=0.001$, $\ell=1$ and $M=1$.}
\begin{tabular}{p{1.2 cm} p{2.8cm} p{3cm} p{2.9cm} p{2.2cm}}
\toprule
$\Lambda$ & \,\,\, $\omega_{max} $ &  \,\,\,\,\,\,$P_{max}$  & \,\,\,\,\,\,\,$P_{tot}$ & \,\,\,\,\,\,\,\,\,\,$\eta$ \\ \midrule
0.01 & 0.2228602 &  $ 1.2032988\times 10^{-7}$ & $1.9319601\times 10^{-8}$ &$409154.25$   \\
0.02 & 0.2111578 & $2.1835963\times 10^{-7}$ & $3.4129539\times 10^{-8}  $ & $207923.76$ \\
0.03 & 0.2005590 &  $3.5567688\times 10^{-7}$ & $5.4154277\times 10^{-8}$  & $118214.79 $
\\
0.04 & 0.1899798 & $5.6078953 \times 10^{-7} $   & $8.3005531\times 10^{-8}$  &  $69203.50$       \\ \bottomrule
\end{tabular}
\end{table}

\begin{eqnarray}\label{h4}
\eta=\frac{\tau_{gap}}{\tau_{emission}},
\end{eqnarray}
where $\tau_{gap}$ represents the average time interval between two successive radiation quanta and $\tau_{emission}$ indicates the time taken for the emission of individual Hawking radiation. Thus, $\tau_{emission}$ is defined as
\begin{eqnarray}\label{h5}
\tau_{gap}=\frac{\omega_{max}}{P_{tot}} \,\, \rm and \,\, \tau_{emission}\geq \tau_{localization}=\frac{2\pi}{\omega_{max}},
\end{eqnarray}
where $\tau_{localization}$ indicates the characteristic time taken of the emitted wave of frequency $\omega_{max}$. The continuous flow of Hawking radiation exists if $\eta\ll 1$ whereas $\eta\gg 1$ indicates a sparse Hawking radiation which also implies that the time gap between the emission of successive quantum radiations is bigger than the time taken for the emission of individual quantum radiation.
The numerical values of $\omega_{max}, P_{max}, P_{tot}$ and $\eta$ for the scalar perturbation are shown in Tables 4, 5 and 6 for different values of parameters $L$, $Q$ and $\Lambda$ respectively. We observe from Tables 4 and 5 that $P_{max}$ and $P_{tot}$ decrease as the values of $L$ and $Q$ increase. However, the sparsity increases as the value of $L$ and $Q$ increase. In Table 6, the values of $P_{max}$ and $P_{tot}$ increase with increasing $\Lambda$ and the sparsity of the black hole decreases with increasing the value of $\Lambda$.

\section{Conclusion}
In this paper, the tunneling of fermion particle for the slowly rotating KNdS-like black hole in bumblebee gravity model is investigated by using modified Dirac equation under the influence of GUP. The Hawking temperature of slowly rotating KNdS-like black hole is modified due to quantum gravity effects. 
It shows that the modified Hawking temperature depends on the mass of the black hole, Lorentz violating parameter, electromagnetic potential, mass and energy of the emitted fermions. If $\beta=0$, the modified Hawking temperature reduces to the original Hawking temperature of slowly rotating KNdS-like black hole. The quantum tunneling rate, modified Hawking temperature and the entropy correction beyond the semiclassical approximation are also derived under GUP. It is noted that the modified Hawking temperature and corrected entropy beyond the semiclassical approximation depend not only on mass of the black hole, Lorentz violating parameter, electromagnetic potential, mass and energy of the emitted fermions but also on Planck constant $\hbar$. The graph of the temperatures $T_0, T_h$ and $T_H$ is plotted in Fig. 2 which obeys the inequality $T_H<T_h<T_0$ and we also observe that $T_h$ coincides with $T_H$ for large value of event horizon.

We present a comprehensive study of  massive scalar perturbation of slowly rotating KNdS-like black hole in bumblebee gravity model. Using the Klein-Gordon equation, we compute the effective potential of massive scalar field. To understand the physical interpretations and the effect of Lorentz violation parameter and other black hole parameters on the effective potentials, we thoroughly investigate the effective potential by plotting it for different values of the parameters. It is found that the peak value of the effective potential decreases with increasing $L$ and $\Lambda$ but the peak value of the effective potential tends to increase with increasing $\ell$ and $Q$. Using the effective potential, we study the greybody factor using the rigorous bound technique. Increasing the values of $L$ tends to increase the greybody factor and this allows the thermal emission of scalar particle more feasible. Moreover, increasing the values of $L$ and $\Lambda$ increase the greybody factor but for varying $\ell$ and $Q$ have the opposite effect. Thus higher effective potential is accompanied with the lower greybody factor and vice-versa.  Our results are also consistent  with quantum mechanics theory. It is noted that for fixed values of overtone number and multipole number, the real parts and the magnitude of imaginary parts of complex frequencies decrease with increasing values of $L$ and $\Lambda$ of the black hole but $Q$ has opposite effect. Therefore, the oscillation frequency and the damping rate decrease with increasing the values of $L$ and $\Lambda$. We also observe that the Hawking sparsity increase with the increasing values of $L$ and $Q$ but decreases with increasing values of $\Lambda$.

For future directions, it would be interesting to study the Dirac, electromagnetic and gravitational perturbation to fully understand the effect of Lorentz invariance violation in the study of gravitational waves. Another aspect of this study is to provide an aspect to determine the exact value of the black hole parameters comparing with the observational data supplied by the EHT for M87* and Sgr A*.
 
{\bf Acknowledgements}: The first author acknowledges the Council of Scientific and Industrial Research, New Delhi, India for giving financial support vide letter No. 211610101868. The authors also thank anonymous reviewers for valuable comments and suggestion to improve the paper.


\begin{thebibliography}{0}
\bibitem{haw2} Hawking, S. W.: Comm. Math. Phys. {\bf 43}, 199 (1975)
\bibitem{page1} Page, D. N.: Phys. Rev. D {\bf 13}, 198 (1976)
\bibitem{page2} Page, D. N.: Phys. Rev. D {\bf 14}, 3260 (1976)
\bibitem{visser} Visser, M.: Phys. Rev. A {\bf 59}, 427 (1999)
\bibitem{boon1} Boonserm, P., Visser, M.: J. Math. Phys. {\bf 51}, 022105 (2010)
\bibitem{fernando} Fernando, S.: Gen. Relativ. Gravit. {\bf 37}, 461 (2005)
\bibitem{kim} Kim, W., Oh, J. J.: J. Korean Phys. Soc. {\bf 52}, 986 (2008)
\bibitem{jusufi} Jusufi, K., Amir, M., Sabir Ali, M., Maharaj, S. D.: Phys. Rev. D {\bf 102}, 064020 (2020)
\bibitem{parikh} Parikh, M. K., Wilczek, F.: Phys. Rev. Lett. {\bf 85}, 5042 (2000)
\bibitem{boon2} Boonserm, P., Ngampitipan, T., Wongjun, P.: Eur. Phys. J. C {\bf 79}, 330 (2019)
\bibitem{sakali} Sakalli, I.: Phys. Rev. D {\bf 94}, 084040 (2016)
\bibitem{badawi} Al-Badawi, A., Sakalli, I., Kanzi, S.: Ann. Phys. {\bf 412}, 168026 (2020)
\bibitem{badawi2} Al-Badawi, A., Kanzi, S., Sakalli, I.: Eur. Phys. J. Plus {\bf 135}, 219 (2020)
\bibitem{gursel} Gursel, H., Sakalli, I.: Eur. Phys. J. C {\bf 80}, 234 (2020)
\bibitem{kanzi1} Kanzi, S., Mazharimousavi, S. H., Sakalli, I.: Ann. Phys.  {\bf 422}, 168301 (2020)
\bibitem{gray} Gray, F., Schuster, S., Van-Brunt, A., Visser, M.: Class. Quant. Grav. {\bf 33}, 115003 (2016)
\bibitem{miao} Miao, Y. G., Xu, Z. M.: Phys. Lett. B {\bf 772}, 542 (2017)
\bibitem{hod1} Hod, S.: Phys. Lett. B {\bf 756}, 133 (2016)
\bibitem{hod2} Hod, S.: Eur. Phys. J. C {\bf 75}, 329 (2015)
\bibitem{chow} Chowdhury, A., Banerjee, N.: Phys. Lett. B {\bf 805}, 135417 (2020)
\bibitem{paul} Paul, A., Majhi, B. R.: Int. J. Mod. Phys. A {\bf 32}, 1750088 (2017)
\bibitem{berti1} Berti, E., Cardoso, V., Starinets, A. O.: Class. Quant. Grav. {\bf 26}, 163001 (2009)
\bibitem{kono1} Konoplya, R. A., Zhidenko, A.: Rev. Mod. Phys. {\bf 83}, 793 (2011)
\bibitem{kono2} Konoplya, R. A.: Phys. Rev. D {\bf 68}, 024018 (2003)
\bibitem{priyo2} Priyobarta, Y. S., Ibungochouba, T. S.: Eur. Phys. J. C {\bf 84}, 1245 (2024)
\bibitem{det} Detweiler, S. l.: Astrophys. J {\bf 239}, 292 (1980)
\bibitem{berti2} Berti, E., Cardoso, V., Will, C. M.: Phys. Rev. D {\bf 73}, 064030 (2006)
\bibitem{pos}  P{\"o}schl, G.,  Teller, E.: Z. Phys. {\bf 83}, 143 (1933)
\bibitem{cho} Cho, H. T.: Phys. Rev. D {\bf 68}, 024003 (2003)
\bibitem{schutz} Schutz, B. F., Will, C. M.: Astrophys. J. {\bf 291}, L33 (1985)
\bibitem{iyer} Iyer, S.: Phys. Rev. D {\bf 35}, 3632 (1987)
\bibitem{zhiden} Zhidenko, A.: Class. Quant. Grav. {\bf 21}, 273 (2004)
\bibitem{cardo1} Cardoso, V., Lemos, J. P.: Phys. Rev. D {\bf 67}, 084020 (2001)
\bibitem{cardo2} Cardoso, V., Natario, J., Schiappa, R.: J. Math. Phys. {\bf 45}, 4698 (2004)
\bibitem{detweiler} Detweiler S.: Astrophys. J. {\bf 239}, 292 (1980)
\bibitem{leaver} Leaver, E. W.: Proc. R. Soc. A {\bf 402}, 285 (1985)
\bibitem{yoshida} Yoshida, S., Futamase, T.: Phys. Rev. D {\bf 69}, 064025 (2004)
\bibitem{chang} Chang, J. F, Shen, Y. G.: Nucl. Phys. B {\bf 712}, 347 (2005) 
\bibitem{van} Vanzo, L., Acquaviva, G., Di Criscienzo, R.: Class. Quant. Grav. {\bf 28}, 183001 (2011)
\bibitem{medov} Akhmedov, E. T., Akhmedova, V., Singleton, D.: Phys. Lett. B {\bf 642}, 124 (2006)
\bibitem{sri} Srinivasan, K., Padmanabhan, T.: Phys. Rev. D {\bf 60}, 024007 (1999)
\bibitem{hemm} Hemming, S., Keski-Vakkuri, E.: Phys. Rev. D {\bf 64}, 044006 (2001)
\bibitem{ame} Amelino-Camelia, G., Arzano, M., Mandanici, G.: Class. Quant. Grav. {\bf 23}, 0264 (2006)
\bibitem{nou} Nouicer, K.: Phys. Lett. B {\bf 646}, 63 (2007)
\bibitem{adler} Adler, R. J., Chen, P. S., Santiago, D. I.: Gen. Relativ. Gravit. {\bf 33}, 1430411 (2001)
\bibitem{fai} Faizal, M., Kruglov, S. I.: Int. J. Mod. Phys. D {\bf 25}, 1650013 (2016)
\bibitem{hao} Zhao, Q., Faizal, M., Zaz, Z.: Phys. Lett. B {\bf 770}, 564 (2017)
\bibitem{gim} Gim, Y., Um, H., Kim, W.: Phys. Lett. B {\bf 784}, 206 (2018)
\bibitem{murei} Mureika, J. R.: Phys. Lett. B {\bf 789}, 88 (2019)
\bibitem{ablu} Ablu, I. M., Ibungochouba, T. S., Gayatri, S. D.: Int. J. Mod. Phys. A {\bf 33}, 1850070 (2018)
\bibitem{kesh} Keshwarjit, A. S., Ablu, I. M., Ibungochouba, T. S., Yugindro, K. S.: Eur. Phys. J. C {\bf 79}, 692 (2019)
\bibitem{priyo} Priyobarta, Y. S., Ibungochouba, T. S., Ablu, I. M., Keshwarjit, A. S.:  Int. J. Mod. Phys. D {\bf 31}, 2250106 (2022)
\bibitem{taw} Tawfik, A., Diab, A.: Int. J. Mod. Phys. D {\bf 23}, 1430025 (2014)
\bibitem{noza} Nozari, K., Azizi, T.: Gen. Relativ. Gravit. {\bf 38}, 0262 (2006)
\bibitem{gan} Gangopadhyay, S., Dutta, A., Saha, A.: Gen. Relativ. Gravit. {\bf 46}, 1661 (2014)
\bibitem{ali} Ali, A.: Phys. Rev. D {\bf 89}, 104040 (2014)
\bibitem{pad} Padmanabhan, T.: Class. Quant. Grav. {\bf 19}, 5387 (2002)
\bibitem{kruglov1} Kruglov, S. I.: Phys. Lett. B {\bf 718}, 15 (2012)
\bibitem{jacob} Jacobson, T., Liberati, I., Crudgington, S., Marringly, D.: Nature {\bf 424}, 1019 (2003)
\bibitem{kruglov2} Kruglov, S. I.: Mod. Phys. Lett. A {\bf 20}, 669 (2003)
\bibitem{ding1} Li, R., Ding, Q. T., Yang, S. Z.: EPL {\bf 138}, 60001 (2022)
\bibitem{liu8}Liu, X., Liu, W., Liu, Z., Wang, J.: arxiv:2503.06404v2  [gr-qc](2025)
\bibitem{liu7} Liu, W., Wen, C., Wang, J.: JHEP {\bf 01}, 184 (2025)
\bibitem{liu6} Liu, W., Fang, X.,Jing, J.,Wang, J.: Sci. China-Phys. Mech. Astron.{\bf 67}, 280413 (2024)
\bibitem{alts} Altschul, B., Bailey, Q. G., Kostelecky, V. A.: Phys. Rev. D {\bf 81}, 065028 (2010)
\bibitem{yang1} Yang, S. Z., Lin, K., Li, J., et al.:  Adv. High Energy Phys. {\bf 2016}, 1 (2016)
\bibitem{yang2} Yang, S. Z., Lin, K.: Acta Phys. Sin. {\bf 68}, 060401 (2019)
\bibitem{yang3} Yang, S. Z., Lin, K.: Acta Phys. Sin. {\bf 190}, 401 (2019)
 \bibitem{onika1} Onika, Y. L., Ibungochouba, T. S.,  Ablu, I. M.: Gen. Relativ. Gravit. {\bf 54}, 77 (2022)
\bibitem{onika2} Onika, Y. L., Ibungochouba, T. S., Ablu, I. M.: Mod. Phys. Lett. A {\bf 38}, 2350089 (2023)
\bibitem{niranjan} Priyobarta, Y. S., Ibungochouba, T. S., Niranjan, S. S.: Chinese Phys. C {\bf 48}, 115111 (2024)
\bibitem{kos1} Kostelecky, V. A., Samuel, S.: Phys. Rev. D {\bf 40}, 1886 (1989)
\bibitem{maluf1} Maluf, R. V., Neves, J. C. S.: Phys. Rev. D {\bf 103}, 044002 (2021)
\bibitem{bluhm1} Bluhm, R., Kostelecky, V. A.: Phys. Rev. D {\bf 71}, 065008 (2005)
\bibitem{maluf2} Maluf, R. V., Almeida, C. A. S., Casana, R., Ferreira, M. M.: Phys. Rev. D {\bf 90}, 025007 (2014)
\bibitem{fang} Liu, W., Fang, X., Jing, J., Wang, J.: Eur. Phys. J. C {\bf 83}, 83 (2023)
\bibitem{uniyal} Uniyal, A., Kanzi, S., Sakall, I.: Eur. Phys. J. C {\bf 83}, 668 (2023)
\bibitem{khoda} Khodadi, M., Schreck, M.: Phys. Dark Universe {\bf 39}, 101170 (2023)
\bibitem{reyes} Reyes, C. M., Schreck, M., Soto, A.: Phys. Rev. D {\bf 106}, 023524 (2022)
\bibitem{casana} Casana, R., Cavalcante, A., Poulis, F. P., Santos, E. B.: Phys. Rev. D {\bf 97}, 104001 (2018)
\bibitem{gullu} Gullu, L., Ovgun, A.: Ann. Phys. {\bf 436}, 168721 (2022)
\bibitem{ovgun1} Ovgun, A., Jusufi, K., Sakalli, I.: Ann. Phys. {\bf 399}, 193 (2018)
\bibitem{ovgun2} Ovgun, A., Jusufi, K., Sakalli, I.: Phys. Rev. D {\bf 99}, 024042 (2019)
\bibitem{ding2} Ding, C., Chen, X., Fu, X.: Nucl. Phys. B {\bf 975}, 115688 (2022)
\bibitem{oli} Oliveira, R., Dantas, D. M., Almeida, C. A. S.: Europhys. Lett. {\bf 135}, 10003 (2021)
\bibitem{ding3} Ding, C., Liu, C., Casana, R., Cavalcate, A.: Eur. Phys. C {\bf 80}, 178 (2020)
\bibitem{jha} Jha, S. K., Rahaman, A.: Eur. Phys. J. C {\bf 81}, 345 (2021)
\bibitem{ding4} Ding, C., Chen, X.: Chin. Phys. C {\bf 45}, 025106 (2021)
\bibitem{maluf3} Maluf, R. V., Muniz, C. R.: Eur. Phys. J. c {\bf 82}, 94 (2022)
\bibitem{wen} Liu, W., Fang, X., Jing, J., Wang, J.: Eur. Phys. J. C {\bf 83}, 83 (2023)
\bibitem{zhou} Liu, J. Z., Guo, W. D., Wei, S. W., Liu, Y. X.: Eur. Phys. J. C {\bf 85}, 145 (2025)

\bibitem{medi} Media, N., Onika, Y. L., Ibungochouba, T. S.: Int. J. Geom. Methods Mod. Phys. {\bf 20}, 2350217 (2023) 
\bibitem{kan} Kanzi, S., Sakalli, I.: Eur. Phys. J. C {\bf 82}, 93 (2022)
\bibitem{ibungo} Ibungochouba, T. S., Ablu, I. M., Yugindro, K. S.: Int. J. Theor. Phys. {\bf 56}, 2640 (2017)
\bibitem{chandra} Chandrashekhar, S.: The Mathematical Theory of Black Holes, Clarendon Press, Oxford (1983)
\bibitem{bonner} Bonner, W., Vaidya, P. C.: Gen. Relativ. Gravit. {\bf 1}, 127 (1970)
\bibitem{chen1} Chen, D. Y., Jiang, Q. Q., Wang, P., Yang, H.:  J. High Energ. Phys. {\bf 11}, 176 (2013)
\bibitem{chen2} Chen, D. Y., Wu, H. W., Yang, H.: JCAP {\bf 03}, 036 (2014)
\bibitem{ran1} Li, R., Yu, Z. H., Yang, S. Z.: Int. J. Theor. Phys. {\bf 62}, 75 (2023)
\bibitem{ran2} Li, R., Ding, Q. T., Yang, S. Z.: EPL  {\bf 138}, 60001 (2022)
\bibitem{ran3} Li, R., Yu, Z. H., Yang, S. Z.: EPL  {\bf 139}, 59001 (2022)
\bibitem{tan1} Tan, X., Zhang, J., Li, R.:  Phys. Scr. {\bf 98}, 105015 (2023)
\bibitem{tan2} Tan, X., Wang, C., Yang, S. Z.: Entropy {\bf 26}, 326 (2024)
\bibitem{ibungo2} Ibungochouba, T. S., Ablu, I. M., Yugindro, K. S.: Astrophys. Space Sci. {\bf 352}, 737 (2014)
\bibitem{media} Media, N., Christina, S., Ibungochouba, T. S.: Int. J. Mod. Phys. A {\bf 39}, 2450105 (2024)
\bibitem{li3} Li, H., Zhang, S., Wu, Y., Zhang, L., Zhao, R.:  Eur. Phys. J. C {\bf 63}, 133 (2009)
\bibitem{ran} Li, R.: Phys. Lett. B {\bf 714}, 337 (2012)
\bibitem{ablu2} Ibungochouba, T. S., Ablu, I. M., Yugindro, K. S.: Int. J. Mod. Phys. D {\bf 25}, 1650061 (2016)
\bibitem{kanzi2} Kanzi, S., Sakalli, I.: Nuclear Phys. B {\bf 946}, 114703 (2019)

\bibitem{ferr} Ferrari, V., Mashhoon, B.: Phys. Rev. D {\bf 30}, 295 (1984)




\end{thebibliography}
\end{document}